\documentclass[epj,nopacs]{svjour}
\usepackage{graphics}
\usepackage{float}
\usepackage{subfigure}
\usepackage{color}
\usepackage{geometry}%
\usepackage{subfig}%
\usepackage{caption}%
\usepackage{amsfonts}
\usepackage{amsmath}
\usepackage{extarrows}
\usepackage{verbatim}
\usepackage{mathrsfs}
\usepackage{booktabs}
\usepackage{verbatim}
\usepackage{epsfig}
\begin{document}
\begin{sloppypar}
\title{Influence of light quark loops on the Wigner phase \textcolor{black}{with Dyson-Schwinger equations approach} }
\author{Jing-Hui Huang\inst{1,2} \thanks{\emph{Email:} jinghuihuang@cug.edu.cn}%
\and Xiang-Yun Hu\inst{1,2}\and  Qi Wang\inst{4} \and  Xue-Ying Duan\inst{3} \and Guang-Jun Wang\inst{3} \and Huan Chen\inst{4}
\thanks{\emph{Email:} huanchen@cug.edu.cn(Corresponding author)}%
}                     
\offprints{}          
\institute{Institute of Geophysics and Geomatics, China University of Geosciences, Wuhan 430074, China \and State Key Laboratory of Geological Processes and Mineral Resources, China University of Geosciences, Wuhan, Hubei, 430074, China \and School of Automation, China University of Geosciences, Wuhan 430074, China \and School of Mathematics and Physics, China University of Geosiciences, Lumo Road 388, 430074 Wuhan, China.}

%
\date{Received: date / Revised version: date}
%
\abstract{
We study the influence of light quark loops on the Wigner phase by solving coupled Dyson-Schwinger equations for quark propagator and gluon propagator. We take the gluon propagator in the Nambu phase from $N_f$ = 2 unquenched lattice QCD and choose various phenomenological models for the quark-gluon vertex. 
\textcolor{black}{
The gluon propagator in Winger phase is assumed to be different from that in the Nambu phase only due to the vacuum polarization of the quark loop. We obtain the Wigner solution of the coupled equations, compared with that from solving only the equation of the quark propagator.} We discussed the corrections by the light quark loops and the dependence on various models of the quark-gluon vertex.
} 
\maketitle
\section{Introduction}	
Due to the nonperturbative character of QCD, the vacuum or ground state of the strong interaction is quite a nontrivial problem. In vacuum, the ground state is characterized by confinement and dynamical chiral symmetry breaking (DCSB), i.e. quarks and gluons are confined in hadrons and light quarks have large constituent masses. Theoretically, there exist also other unstable or metastable states (phases), which are represented as multi-solutions of the equation of motion of QCD. Usually, the phase with DCSB is called the Nambu phase, while the phase with chiral symmetry restored is called the Wigner phase. It is also very important to investigate these unstable or metastable states.  \textcolor{black}{It helps us to understand the symmetry~\cite{Klahn2016,QCDPT-DSE12-4,Li:2018tut}, vacuum energy~\cite{Barnafoldi2007}, etc.
More importantly, due to asymptotic freedom in the ultraviolet region, one could expect phase transition with deconfinement and chiral symmetry restoration at high temperature and/or chemical potentials. }

The phase diagram of strong-interaction matter has been studied since the early days of quantum field theory~\cite{Cornwall1974,bag2,bag4}. And it is a hot topic in recent nuclear physics study~\cite{Masuda2014,Masuda2015,Masuda2016,DElia2019,PhysRevD.99.014007}. A smooth crossover at low chemical potential and high temperature has been widely accepted due to \textcolor{black}{ heavy-ion collision experiments~\cite{PhysRevLett.127.262301} and various theoretical studies in recent years\cite{PhysRevD.94.054008,FISCHER2011438,PhysRevD.101.054032,Golanbari_2020,Bazavov2019,Guenther2021}}. While at non-vanishing chemical potential and lower temperature, first-order phase transitions may take place in the beam energy scan experiments of STAR at BNL, the core of massive neutron stars, the collision and \textcolor{black}{coalescence of binary neutron stars~\cite{PhysRevD.100.024061}}, etc. 


\textcolor{black}{Theoretically, the Dyson-Schwinger Equations (DSEs) of QCD are the equations of motion in the continuum quantum field theory, which provide a proper framework to investigate unstable or metastable states and the quark matter with high baryon chemical potentials in absence of "sign problem". They are a set of coupled equations of quark, gluon, ghost propagators, and higher point Green functions, and in general a practical truncation scheme is needed to get a solvable system.  Based on successful phenomenological descriptions of hadrons in vacuum (see, e.g., Refs.~\cite{ROBERTS2021103883,Roberts-Hadron-7,EICHMANN2012234,CLOET20141,HUBER20201})}, they are extended to study phase transitions in hot/dense medium (see, e.g., Refs.~\cite{QCDPT-DSE11,QCDPT-DSE12-5,QCDPT-DSE23}), and even the equation of state of dense quark matter in compact stars~\cite{Chen2015,PhysRevD.103.103003,Bai2021}. Among these work, the Wigner phase is a key element. However, in the computation of the Wigner phase, models of gluon propagator and quark-gluon vertex are usually taken \textcolor{black}{ from Lattice QCD~\cite{Sternbeck2010,PhysRevD.86.114513,PhysRevD.102.114518}} or by fitting hadron properties  ~\cite{DSE-1-1,DSE-1-2,DSE-1-3}, which are all set in the Nambu phase. The feedback of quarks in the Wigner phase is usually neglected. It is a question of whether they are good approximations in describing the Wigner phase and how large the corrections are. Especially in the case of dense quark matter, the feedback of quarks to gluons would be important.

\textcolor{black}{In this work, we intend to investigate such feedback effects in the Wigner phase with baryon chemical potential $\mu_B=0$ as the first step. We will improve the truncation scheme of the DSEs, focusing on the influence of the quark loops in the vacuum polarization of the gluon propagator. A coupled DSEs of quark and gluon propagators in the Wigner phase will be solved to investigate the corrections.}

The paper is organized as follows. In section 2, the truncation scheme of DSE for quark propagator and gluon propagator in the Nambu and Wigner phase is given. 
In section 3,  the numerical results of the quark and gluon propagator in the wigner phase are given and discussed. Finally, we summarize our work and give a brief remark in section 4.

\section{Truncation scheme for DSEs in Nambu and Wigner phase}

\subsection{Nambu Phase}

\begin{figure}[htp!]
 \centering
 \includegraphics[scale=0.7]{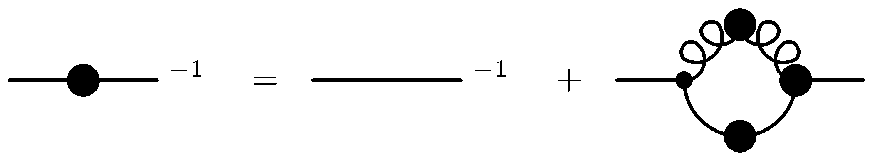}\\[1.5em]
 \includegraphics[scale=0.5]{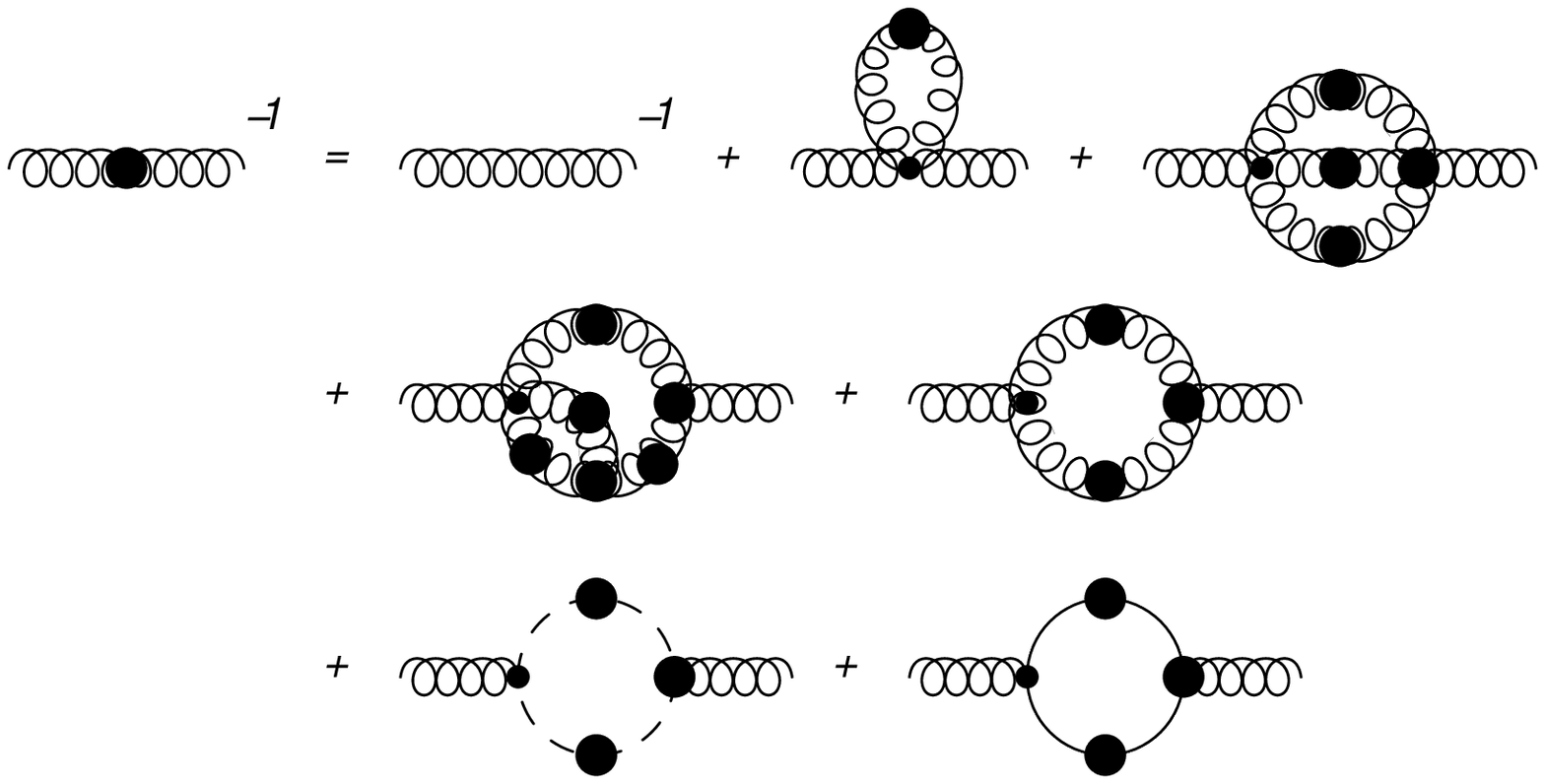}
 \caption{\label{Quark_Gluon_DSE}%
 Dyson-Schwinger equation for quark propagator and gluon propagator.}
\end{figure}

The basic degrees of freedom in QCD are quarks and gluons. The DSEs of the quark propagator and gluon propagator is depicted in Fig.~\ref{Quark_Gluon_DSE}.
The quark propagator and gluon propagator depend on the ghost propagator and various vertices, 
which depend on higher-order Green functions in their own DSEs.
To get a solvable system, one must truncate and close the system.

To investigate the Nambu phase with DCSB and the Wigner phase, one first focus on the quarks. The DSE of quark propagator in vacuum  is expressed in Euclidean space as
\begin{flalign}
\label{Eq:quarkDSEdefine}
&\ S(p)^{-1}= Z_2 (i\gamma\cdot \tilde{p}+m_{q})+
  \Sigma(p), &\\ 
&\ \Sigma(p)= Z_1 g^2\int \frac{d^4q}{(2\pi)^4}D_{\rho\sigma}(k)\delta_{ab} \frac{\lambda^a}{2}\gamma_\rho S(q)
\frac{\lambda^b}{2}\Gamma_\sigma(q,p)&,
\nonumber
\end{flalign}
where $k=p-q$,  trivial color structures are taken for the full gluon propagator $D_{\rho\sigma}(k)\delta_{ab}$
and full quark-gluon vertex $ \frac{\lambda^b}{2}\Gamma_\sigma(q,p)$,
$Z_{1}$ is the renormalization constant for quark-gluon vertex, $Z_{2} $ is the quark wave-function renormalization constant, \textcolor{black}{$m_q$ represents current quark mass. In this work, we investigate only the light quarks (u and d) and set the current quark mass  ($m_{q}=0$), i.e. the chiral limit since it is quite small comparing with the dynamical quark mass in the Nambu phase and the baryon chemical potential in dense quark matter.
The chiral limit approximation for ligh quarks has been widely used for investigating hadron properties~\cite{ROBERTS2021103883,DSE-1-1,DSE-1-2,DSE-1-3} and phase transitions~\cite{HE200932,YUAN200669,Chen2015,PhysRevD.103.103003}
}

The general structure of the quark propagator in vacuum can be expressed with the Lorentz covariant form
\begin{flalign}
&\ S^{-1}(p)=i{\gamma}\cdot
{p}A({p}^2)+ B({p}^2), &
\end{flalign}
where $A({p}^2)$ and $B({p}^2)$ are scalar functions of $p^2$.
In chiral limit or for light quarks, one could obtain two solutions for the quark propagator.
One solution is characterized by nonzero dynamical mass $ M(p^2)=\frac{B({p}^2)}{A(p^2} \ne 0$ in chiral limit.
which is called the Nambu solution (phase).
The other solution is characterized by $B({p}^2) = 0$ in the chiral limit, which is called the Wigner solution (phase).
Of course, the real physical vacuum is the Nambu phase while the Wigner phase is a metastable or unstable phase.

\begin{figure*}[htp!]
	\centering
\subfigure
{
	\vspace{-0.2cm}
	\begin{minipage}{8.2cm}
	\centering
	\centerline{\includegraphics[scale=0.652,angle=0]{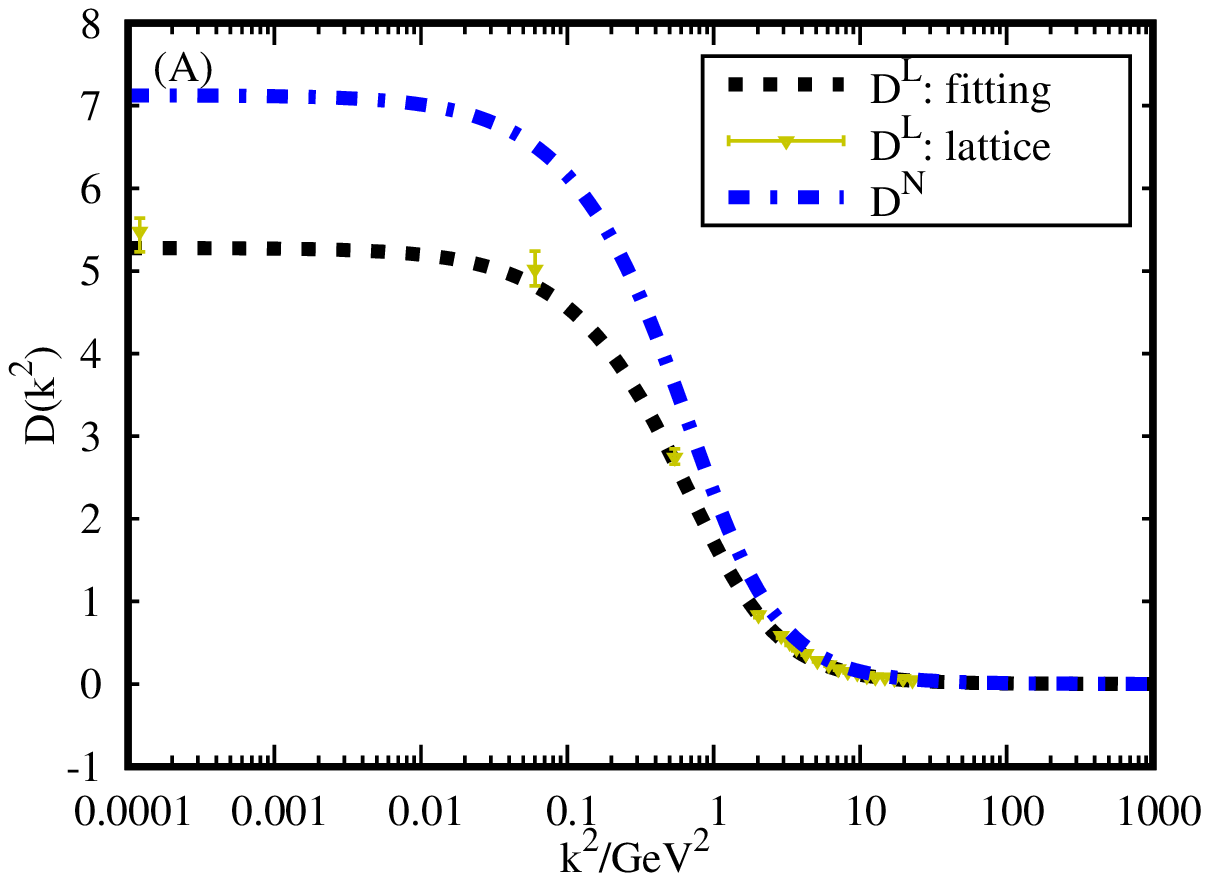}}
	\end{minipage}
}
\vspace{-0.2cm}
\subfigure
{
	 \vspace{-0.2cm}
	\begin{minipage}{8.2cm}
	\centering
	\centerline{\includegraphics[scale=0.652,angle=0]{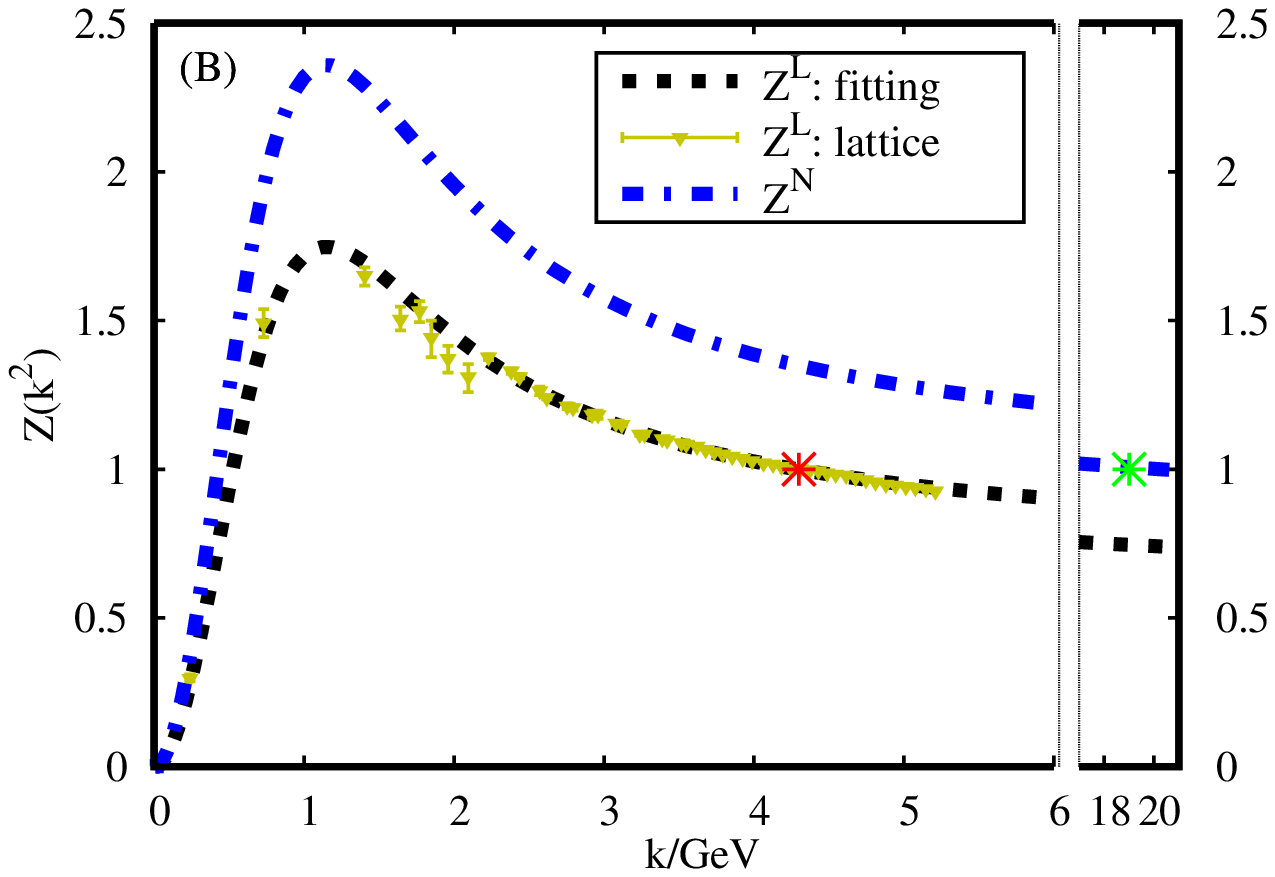}}
	\end{minipage}
}
\vspace*{0mm} \caption{\label{InputWinger_latticePropagator}
\textcolor{black}{
The gluon propagator function $D(k^2)$ (left panel) and the gluon dressing function $Z(k^2)$ (right panel). The superscript $L$ represents the results from the lattice QCD~\cite{Ayala:2012pb} , which are renormalized with $\xi =4.3 {\rm GeV}$.
The superscript $N$ represents the gluon propagator used in the DSE for studying the Nambu phase with renormalized point $\xi =19 {\rm GeV}$. The stars represent corresponding renormalization points. }
}
\end{figure*}

To solve the DSE for the quark propagator, one needs the full gluon propagator and quark-gluon vertex. The full gluon propagator is studied with the DSE approach~\cite{Papavassiliou:2014qva,Meyers:2014iwa,Lowdon:2018mbn} and the lattice QCD~\cite{Ayala:2012pb,PhysRevD.102.114518,PhysRevD.98.014002}. The results with the DSE approach still depend on the various ansatz of quark-gluon vertex and higher-point green functions\cite{PhysRevD.81.065003,PhysRevD.100.056001}. In the following, we adopt the gluon propagator in the Nambu phase from the Lattice QCD\cite{Ayala:2012pb}. The gluon propagator in the Landau gauge is expressed as
\begin{flalign}
&\  D_{\rho \sigma}(k)  =   D(k^2)T_{\rho\sigma}(k) =Z(k^2)D^0_{\rho\sigma}(k) \, ,&
\label{Nambu_gluon}
\end{flalign}
where $T_{\rho\sigma}(k)=\Big[\delta_{\rho\sigma}-\frac{k_\rho k_\sigma}{k^2} \Big]$, $D^0_{\rho\sigma}(k)=\frac{1}{k^2}T_{\rho\sigma}(k)$ is the free gluon propagator, the scalar functions $D(k^2)$, $Z(k^2)$ are called the gluon propagator function and gluon dressing function respectively. 
\textcolor{black}{The data of the gluon propagator from the unquenched lattice QCD with $N_f=2$ and lattice volume v=$48^{3} \times 96$ ~\cite{Ayala:2012pb} are shown in Fig.~\ref{InputWinger_latticePropagator}. For convenience we fit the data with the form as in Ref.~\cite{DUDAL2018351,Oliveira2019}:
}
\textcolor{black}{
\begin{flalign}
\label{Eq:fitting-gluonLattice}
&\ 
D^{L}(k^{2})= \frac{Z_0(k^{2}+M_{1}^{2})}{k^{4}+M_{2}^{2}k^{2}+M_{3}^{4}} \left [ 
\overline{ \omega} {\rm ln} \left ( \frac{k^{2}+m_{0}^{2}}{\Lambda_{QCD}^{2}} \right )+1
\right ]^{\gamma_g},
&\
\end{flalign}
where $Z_0$=1.3548, $M_{1}^{2}$ = 4.2 ${\rm GeV^{2}}$, $M_{2}^{2}$ = 1.2121 ${\rm GeV^{2}}$, $M_{3}^{4}$ = 1.05 ${\rm GeV^{4}}$, $m_{0}^{2}$ = 0.216 ${\rm GeV^{2}}$, $\Lambda_{QCD}$ = 0.425 ${\rm GeV}$, $\beta_{0}=(11N_{c}-2N_{f})/3$, $\overline{ \omega}= \beta_{0} \alpha(\xi)/(4 \pi)$
with a strong coupling constant $\alpha(\xi =4.3~{\rm GeV})=0.38$, and the gluon anomalous dimension  $\gamma_g=\frac{4N_f-13N_c}{22N_c-4N_f}$. 
Note herein and in the following, we will take all $N_f=2$ and $\Lambda_{QCD}$ = 0.425 ${\rm GeV}$ for self-consistence, which are different from Ref.~\cite{DUDAL2018351,Oliveira2019}. 
This form can be extended to ultraviolet and gives the consistent perturbative one-loop behavior of the gluon propagator.
Though finite quark masses are adopted in lattice QCD calculations~\cite{Ayala:2012pb}, the effects are also neglected here concerning the large dynamical quark masses. 
One should also note that the renormalization point $\xi$ =$4.3~{\rm GeV}$ in the lattice QCD calculations for the gluon propagator is different from the renormalization point $\xi$ =$19~{\rm GeV}$ for solving the DSEs in the following.
Therefore, we further multiply the Lattice's gluon function $Z^{L}$, $D^{L}$ by an additional renormalization constant $R_{L}$:
\begin{flalign}
\ \label{ZN} 
&\  Z^{N}(k^2)=R_{L} Z^{L}(k^2), &\
\end{flalign}
\begin{flalign}
\label{DN}
&\  D^{N}(k^2)=R_{L} D^{L}(k^2), &\  
\end{flalign}
with the renormalization constant $R_{L} = 1.35$ fixed by setting $Z^{N}(\xi =19 ~{\rm GeV})$ = 1, the superscripts $N$ representing corresponding results in the Nambu phase. The lattice results and our renormalized results of the gluon propagator function and gluon dressing function are shown in Fig.~\ref{InputWinger_latticePropagator}.}

\textcolor{black}{The quark-gluon vertex is much more complicated, which includes 12 independent Lorentz matrix structures. Though it is investigated with lattice QCD~\cite{PhysRevD.103.114515}, DSE approach~\cite{Williams2015,PhysRevD.100.056001} and functional renormalization group~\cite{PhysRevD.97.054006}, there are still many uncertainties to determine the full quark-gluon vertex. 
Phenomenological models for the quark-gluon vertex are still widely used in the literature~\cite{ROBERTS2021103883,Roberts-Hadron-7,EICHMANN2012234,CLOET20141,HUBER20201}). The ansatz for solving Eq.(\ref{Eq:quarkDSEdefine}) is often taken as
\begin{flalign}
\label{KernelAnsatz}
&\ { Z_{1}g^2 D_{\rho \sigma}(k) \Gamma_\sigma(q,p)}  \nonumber &  \\ 
&\ \quad \quad \quad   =  [Z_{2}]^{2} g^2 [Z^N(k^{2}) D^0_{\rho \sigma}(k)] [{\Gamma}(k^2)\tilde{\Gamma}_{\sigma}(q,p)] & \\
&\ \quad \quad \quad   = [Z_{2}]^{2} {\cal G}(k^2)D^0_{\rho \sigma}(k){\tilde{\Gamma}}_{\sigma}(q,p) \, .\nonumber & 
\end{flalign}
}
\textcolor{black}{where the quark-gluon vertex is separated into two parts with some arbitrariness, $\tilde{\Gamma}_{\sigma}(q,p)$ represents the matrix structure parts of the quark-gluon vertex, ${\Gamma}(k^2)$ represents the remaining part of the vertex and depends only on the gluon momentum $k^2$. In the following, we call $\tilde{\Gamma}_{\sigma}(q,p)$ the vertex structure and $g^2{\Gamma}(k^2)$ the vertex dressing function, respectively.
The so called effective interaction ${\cal G}(k^2)=g^2Z^N_{}(k^{2}){\Gamma}(k^2)$ includes both contributions from the gluon propagator and quark-gluon vertex. Consequently, the vertex dressing function $g^2{\Gamma}(k^2)$ can be obtained from the gluon dressing function and the effective interaction ${\cal G}(k^2)$ 
\begin{flalign}
\label{vertex}
&\ g^{2} {\Gamma}^{}(k^2)= \frac{{\cal G}(k^2)}{ D^N(k^2) k^2}= \frac{{\cal G}(k^2)}{ Z^N(k^2) } \, .&
\end{flalign}
}

\textcolor{black}{
In the following we consider two widely used models for the effective interaction ${\cal G}(k^2)$, the renormalization-group-improved Maris-Tanday(MT) model~\cite{Maris1999} and Qin-Chang(QC) model~\cite{Qin2011}, %
\begin{flalign}
\label{MTg}
&\ \frac{{\cal G}^{MT}(k^2)}{k^2}= \frac{4\pi^2}{\omega^6} \, \hat{D}\,k^2\, {\rm e}^{-k^2/\omega^2}+ \mathcal{F}(k^2)\, ,&
\end{flalign}
\begin{flalign}
\label{QCg}
&\ \frac{{\cal G}^{QC}(k^2)}{k^2}= \frac{8\pi^2}{\omega^4} \, \hat{D} {\rm e}^{-k^2/\omega^2}+ \mathcal{F}(k^2)\, , &
\end{flalign}
\begin{flalign}
\label{weirao}
&\ \mathcal{F}(k^2)=  \frac{8 \pi^{2} \gamma (1-e^{-k^{2}/4m_{t}^2})}{\ln[\tau +(1+k^{2}/\Lambda _{QCD}^{2})^2]}, &
\end{flalign}
where $\tau =e^{2} -1$, $m_t$ = 0.5 GeV, the anomalous dimension $\gamma=12/(11N_c-2N_f)$,
$N_f=2$ and $\Lambda _{QCD}^{}$ = 0.425 GeV are the same as in the gluon propagator Eq.\ref{Eq:fitting-gluonLattice}, 
the parameter $\hat{D}$ and $\omega$ in the infrared part are usually fixed by fitting meson properties obtained by solving consistent Bethe-Salpeter equations~\cite{PhysRevC.84.042202,PhysRevD.103.074001}. Eq.~(\ref{weirao}) corresponds to the strong fine-structure constant at the renormalization scale 
\textcolor{black}{
$\alpha(\xi=19GeV)=0.216$.} The corresponding quark-gluon vertex dressing functions are remarked as ${\Gamma}^{MT}(k)$ and ${\Gamma}^{QC}(k^2)$ respectively. 
}

\textcolor{black}{
A similar truncation scheme is also investigated in Refs.~\cite{PhysRevD.80.074029,Fischer2010fx,Fischer2012vc,PhysRevD.101.014016}, in which one inputs a fitting form of the gluon propagator with quenched ($N_f=0$) lattice QCD, and a corresponding quark-gluon vertex ansatz. Then coupled DSEs of the quark propagator and gluon propagator are solved to investigate the unquenching effects on the quark propagator and gluon propagator. It is different from our truncation scheme, which starts directly from the unquenched gluon propagator and quark-gluon vertex in the Nambu phase. In the following, to study the model dependence of our results, we would also employ an "CF model", i.e. a vertex dressing function similar as in Ref.~\cite{PhysRevD.80.074029,Fischer2010fx,Fischer2012vc,PhysRevD.101.014016}
\begin{flalign}
g^2\Gamma^{CF} \left( k^{2} \right) = &4\pi\alpha(\xi) \tilde{R} \left[ \frac{d_{1}}{d_{2}+k^{2}} +\frac{k^{2}}{\Lambda_{QCD}^{2}+k^{2}} \right.
&  \nonumber\\ 
& \left. \times \left(\frac{\beta_{0} \alpha(\xi) \ln \left[k^{2} / \Lambda_{QCD}^{2}+1\right]}{4 \pi}\right)^{2 \delta} \right], & 
\label{Eq.lattice_vertex}
\end{flalign}
where the parameters $d_{1}$, $d_{2}$ in the first term are relevant for the infrared behaviour. The second term in parentheses describes the perturbative running in the ultraviolet, with the anomalous dimension $\delta =-9N_{c}/(44N_{c}-8N_{f})$. The strong fine-structure constant at the renormalization scale was taken as $\alpha(\xi=4.47\, {\rm GeV})=0.3$ ~\cite{Muller2013,PhysRevD.80.074029}. However, we would take consistent parameters as in Eq.(\ref{Eq:fitting-gluonLattice}), which are different from the quenched approximation $N_f=0$. In addition, we also shift the renormalization scale to $\xi=19GeV$ with an additional factor $\tilde{R}=0.74$. The parameters $d_{1}$, $d_{2}$ would be set in next section.}



There are several widely used Phenomenological models for the structure of quark-gluon vertex, e.g. the rainbow (RB) approximation, the Ball-Chiu (BC) vertex~\cite{BC19811,Oliveira2019} ans\"{a}tz form, 
\textcolor{black}{
or even more complicated forms~\cite{PhysRevD.99.074013,PhysRevD.100.056001}.
}
In the following, we investigate three models, the rainbow approximation, the BC vertex ans\"{a}tz, and the BC1 form.
In the rainbow approximation, the bare vertex is used for the structure of the quark-gluon vertex
\begin{flalign}
\label{rainbowvetex} 
&\ \tilde{\Gamma}^{RB}_\sigma(q,p)  = \gamma_{\sigma} \, . &\
\end{flalign}
The form of the BC vertex is developed in \cite{BC19811,BC1981},
\begin{flalign}
&\ \tilde {\Gamma}_\sigma^{BC}(q,p) = \lambda_{1} \gamma_{\mu}
+\lambda_{2} (p+q)_{\mu}  \nonumber\\
&\ \qquad \qquad \quad +\lambda_{3} (p+q)_{\mu}(p\cdot\gamma+q\cdot\gamma) .& 
\label{bcvtxmu}
\end{flalign}
where $\lambda_{i}$(i=1,2,3) are functions of $A(p^{2})$, $A(q^{2})$, $B(p^{2})$ and $B(q^{2})$:
\textcolor{black}{
\begin{flalign}\label{2}
&\ \lambda_{1}=\frac{1}{2}\frac{A(p^{2})+A(q^{2})}{2},\nonumber  & \\
&\ \lambda_{2}=-i\frac{B(p^{2})-B(q^{2})}{p^{2}-q^{2}}, & \\
&\ \lambda_{3}=\frac{1}{2}\frac{A(p^{2})-A(q^{2})}{p^{2}-q^{2}}. \nonumber &
\end{flalign}
}
As a comparison, we also investigate the BC1 approximation, i.e. the first item of the BC vertex:
\begin{flalign}
&\ \tilde {\Gamma}_\sigma^{BC1}(q,p) =  \lambda_{1} \gamma_{\sigma} .  &
\label{bconevtxmu}
\end{flalign}

With all these above inputs, one can now solve the DSE for the quark propagator Eq.~(\ref{Eq:quarkDSEdefine}) in Nambu phase, and further study hadron properties in combination with Bethe-Salpeter equations for meson and Faddeev equations for baryons.


\subsection{Winger Phase}
\begin{figure}[htp!]
 \centering
 \includegraphics[scale=0.61]{qdse.eps}\\[1.5em]
 \includegraphics[scale=0.61]{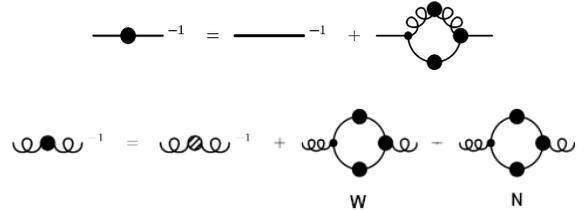}
 \caption{\label{traunc_Quark_Gluon_DSE}%
Our truncation scheme for the coupled DSEs of quark propagator and gluon propagator in the Wigner phase. The shaded gluon propagator is the gluon propagator in Nambu phase given by Eq. (\ref{DN}). ’W’/'N' represents calculation with quark propagator in Wigner phase and Nambu phase respectively.}
\end{figure}

To investigate the Wigner phase, one usually keeps the ansatz Eq.~(\ref{KernelAnsatz}) and only solves the DSE for the quark propagator by setting $B(p^2)=0$ in the chiral limit\cite{DSE-1-2,Zong2002,Chen2008}. 
\textcolor{black}{In the following, we call this truncation scheme the old truncation scheme.}
However, one should note that the gluon propagator, the quark-gluon vertex
and the effective interaction ${\cal G}(k^2)$ in the above section are obtained in the Nambu phase, which should be all modified in the Wigner phase. 
\textcolor{black}{As an obvious modification, one can see the direct feedback of the quark propagator to the gluon propagator via the quark loop in Fig.~\ref{Quark_Gluon_DSE}}.
To investigate such an modification, for the first step, we assume the difference of the gluon propagator in the Nambu phase and Wigner phase is only contributed by the quark loop, neglecting further modifications in other diagrams. Therefore, we have
\begin{flalign}
\label{gdse0}
&\  (D^W)^{-1}_{\mu\nu}(k) - \Pi^{W}_{\mu\nu}(k)= (D^N)^{-1}_{\mu\nu}(k)-\Pi^{N}_{\mu\nu}(k), &
\end{flalign}
where the superscripts $W,N$ represent calculations in Wigner phase and Nambu phase respectively, and the vacuum polarization tensor $\Pi^{W,N}_{\mu\nu}$ are calculated from only the quark loop diagram, 
\begin{flalign}
  \label{polarization}
&\ \delta_{ab}\Pi_{\mu\nu}(k) = -{N_{f}}\int_{q} {\rm Tr}[\gamma_{\mu} \frac{\lambda^a}{2} S(p) \nonumber &\ \\
&\ \quad \quad \quad \quad \quad \quad 
\times g^2{\Gamma}(k^2)\tilde{\Gamma}_{\nu}(q,p) \frac{\lambda^b}{2} S(q)] \, ,&
\end{flalign}
with $p = k + q$, and '{\rm Tr}' in Dirac space.
Therefore, we obtain the coupled DSEs for quark propagator and gluon propagator in the Wigner phase
\begin{flalign}
\label{Eq:qdseInWigner1}
&\ S(p)^{-1}= Z_2 (i\gamma\cdot \tilde{p}+m_{q}) +\textcolor{black}{ \Sigma^{W} (p)}, &\ 
\end{flalign}
\begin{flalign}
\label{Eq:qdseInWigner2}
&\  (D^W)^{-1}_{\mu\nu}(k) = (D^N)^{-1}_{\mu\nu}(k) + \Pi^{W}_{\mu\nu}(k)-\Pi^{N}_{\mu\nu}(k). &\
\end{flalign}
\textcolor{black}{Our new truncation scheme is depicted in Fig.~\ref{traunc_Quark_Gluon_DSE}.
In this work, we neglect further corrections on the quark-gluon vertex except the matrix structure in BC vertex and BC1 vertex, which depends explicitly on the quark propagator.}

To keep the transversality of the gluon propagator, we retain only the transverse part of the polarization tensor $\Pi_{\mu\nu}(k)$, i.e.
\begin{flalign}
\label{polarization-cast}
&\ \Pi(k^2)  = \frac{1}{3}T_{\mu\nu} (k)\Pi_{\mu\nu}(k)\, , &  \\
&\ \Pi_{\mu\nu}(k) = \Pi(k^2) T_{\mu\nu} (k)\, , &
\end{flalign}
where the term $\Pi(k^2)$ is called the polarization scalar function. We then obtain the expression for $\Pi_i(k^2)$ from Eq.~(\ref{polarization}) with subscript $i=1,2,3$ corresponding to results with the bare vertex, the BC1 vertex, and the BC vertex respectively,
\begin{flalign}
\label{polarization211}
&\ \Pi_{1}(k^2)=  {\cal H} (k^2)\int_{q} \frac{dq^4}{\Delta (p^{2})\Delta (q^{2})} (I_{1}+I_{2}),   & 
\end{flalign}
\begin{flalign}
\label{polarization212}
&\ \Pi_{2}(k^2)=  {\cal H} (k^2)\int_{q} \frac{dq^4}{\Delta (p^{2})\Delta (q^{2})} (I_{3}+I_{4}),  & 
\end{flalign}
\begin{flalign}
\label{polarization213}
&\ \Pi_{3}(k^2)=  {\cal H} (k^2)\int_{q} \frac{dq^4}{\Delta (p^{2})\Delta (q^{2})} (I_{5}+I_{6}+I_{7}),  &\
\end{flalign}
with
\begin{flalign}
&\ {\cal H} (k^2)= -\frac{4g^{2}\Gamma(k^2)}{6 \times (2\pi)^{4}}, \nonumber &\ \\ 
&\  \Delta (p^{2})=p^{2} A^{2}(p^2)+B^{2}(p^2), \nonumber  &\  \\
&\ \Delta (q^{2})=q^{2} A^{2}(q^2)+B^{2}(q^2),\nonumber &\
\end{flalign}
\textcolor{black}{
and the integral kernels $I_{1}$, $I_{2}$, $I_{3}$, $I_{4}$, $I_{5}$, $I_{6}$ and $I_{7}$ are defined as
}
\begin{flalign}
&\ I_{1}=3B(p^2)B(q^2), \nonumber  &\ \\ 
&\ \textcolor{black}{ I_{2}=(p\cdot q+2 \frac{k\cdot p \times k\cdot q}{k^2}) A(p^2)A(q^2),} \nonumber &\ \\
&\ I_{3}=\lambda_{1} B(p^2)B(q^2), \nonumber &\ \\ 
&\ I_{4}=2(p\cdot q) \lambda_{1}  A(p^2)A(q^2), \nonumber &\ \\
&\ I_{5}=(\lambda_{1}+t\cdot q\lambda_{3}+t\cdot p\lambda_{3}) B(p^2)B(q^2), \nonumber &\ \\ 
&\  I_{6}=-i\lambda_{2}(p\cdot t)B(q^2)A(p^2)-i\lambda_{2}(q\cdot t)B(p^2)A(q^2), \nonumber&\  \\ 
&\  I_{7}=(2(p\cdot q)\lambda_{1}-(t\cdot q)(p\cdot q)\lambda_{3}-(t\cdot p)q^{2}\lambda_{3})A(p^2)A(q^2). \nonumber  &\
\end{flalign}
\textcolor{black}{where $t= p+q$.
Finally, we obtain the full gluon propagator function $D^{W}_{}(k)$ in the Wigner phase
\begin{flalign}
\label{polarization-p3p42}
&\ D^{W}_{}(k) =\frac{D^N(k)}{1 + D^N(k)(
    \textcolor{black}{\Pi^{W}_{}(k)-\Pi^{N}_{}(k)}
    )} \,. &\
\end{flalign}}

\textcolor{black}{
Note that though the integrals in Eqs.~(\ref{polarization211}),(\ref{polarization212}),(\ref{polarization213}) are superficially  (quadratic) divergent, we only need to calculate the difference of the polarization scalar function of the two phases $\Pi^{W}(k)-\Pi^{N}(k)$ in our scheme. Since the Nambu solution and the Wigner solution coincide highly in ultraviolet, the integral with the subtraction is finite and we do not need further renormalization.}

\begin{table*}[htp!]
\renewcommand\arraystretch{1.5}
\centering
\caption{\label{tab:table4}%
\textcolor{black}{
Model and parameter settings as well as some characteristic numerical results with the vertex dressing function ${{\Gamma}^{MT}(k^2)}$ and ${{\Gamma}^{QC}(k^2)}$.
}
}
\vspace{-0.25cm}
\begin{tabular}{l@{\hspace{1.01cm}}l@{\hspace{1.01cm}}l@{\hspace{1.01cm}}l@{\hspace{0.7cm}}l@{\hspace{0.7cm}}l@{\hspace{0.7cm}}c}
\hline
\hline
model &
${\Gamma}(k^2)$&
$\tilde{\Gamma}_{\sigma}(q,p)$& 
$\omega $ (GeV)& 
$\hat{D} ~{\rm (GeV^{2})}$   & $-\langle \bar qq \rangle^{1/3}_{0}$~ (MeV) &
$(\omega \hat{D})^{1/3} ~{\rm (GeV)}$\\
\hline
DSE1  &MT &RB&   0.60 &  0.85 & 280   & 0.82    \\
DSE2  &MT &BC1 &  0.60 &  0.64 & 279   & 0.74 \\
DSE3  &MT &BC&     0.60 &  0.44 & 281   & 0.65  \\
DSE4  &QC &RB&     0.65 &  0.91 & 281   & 0.85     \\
DSE5   &QC &BC1&    0.70 &  0.62 & 281   & 0.77  \\
DSE6   &QC &BC&     0.70 &  0.35 & 282   & 0.65\\
\hline
DSE41   &QC &RB&   0.65 &  0.84 & 271   & 0.83    \\
DSE42   &QC &RB&   0.65 &  0.99 & 290   & 0.87    \\
DSE61   &QC &BC&   0.70 &  0.33 & 268   & 0.62\\
DSE62   &QC &BC&   0.70 &  0.37 & 291   & 0.66\\
 \hline
\hline
\end{tabular}
\end{table*}
\begin{table*}[htp!]
\renewcommand\arraystretch{1.5}
\centering
\caption{\label{tab:table5}%
\textcolor{black}{
Model and parameter settings as well as some characteristic numerical results with the vertex dressing function ${{\Gamma}^{CF}(k^2)}$.}}
\vspace{-0.25cm}
\begin{tabular}{l@{\hspace{1.25cm}}l@{\hspace{1.21cm}}l@{\hspace{1.25cm}}l@{\hspace{1.35cm}}l@{\hspace{1.4cm}}l@{\hspace{1.4cm}}c}
\hline
\hline
model &
${\Gamma}(k^2)$&
$\tilde{\Gamma}_{\sigma}(q,p)$& 
$d_1 $ (GeV)& 
$d_2 $ $(\rm GeV^{2})$  & $-\langle \bar qq \rangle^{1/3}_{0}$~ (MeV) \\
\hline
DSE7  &CF &RB  &   6.10 &  0.5 & 280       \\
DSE8  &CF &BC1 &   4.61 &  0.5 & 279    \\
DSE9  &CF &BC  &   2.65 &  0.5 & 280    \\
 \hline
\hline
\end{tabular}
\end{table*}
\begin{figure*}[htp!]
	\centering
\subfigure
{
	\vspace{-0.2cm}
	\begin{minipage}{8.2cm}
	\centering
	\centerline{\includegraphics[scale=0.652,angle=0]{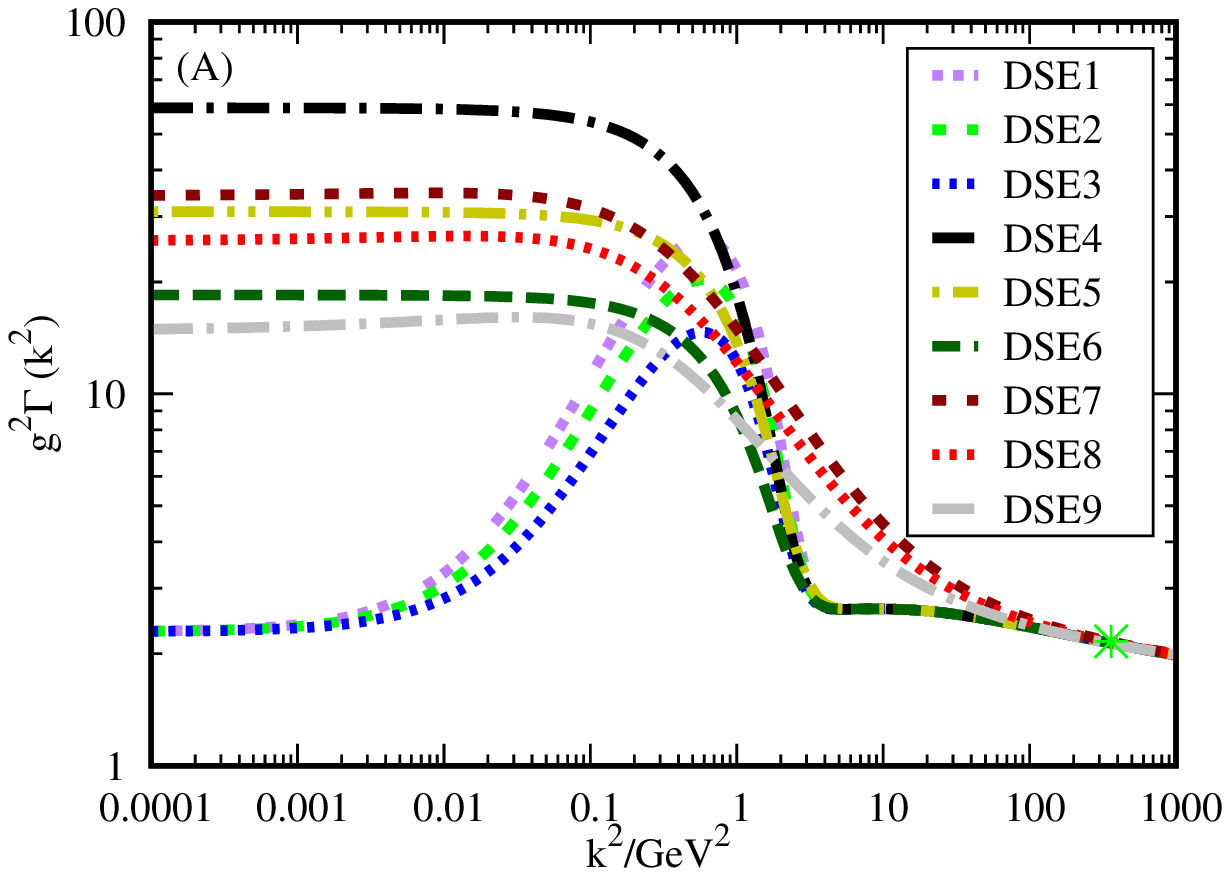}}
	\end{minipage}
}
\vspace{-0.2cm}
\subfigure
{
	\vspace{-0.2cm}
	\begin{minipage}{8.2cm}
	\centering
	\centerline{\includegraphics[scale=0.652,angle=0]{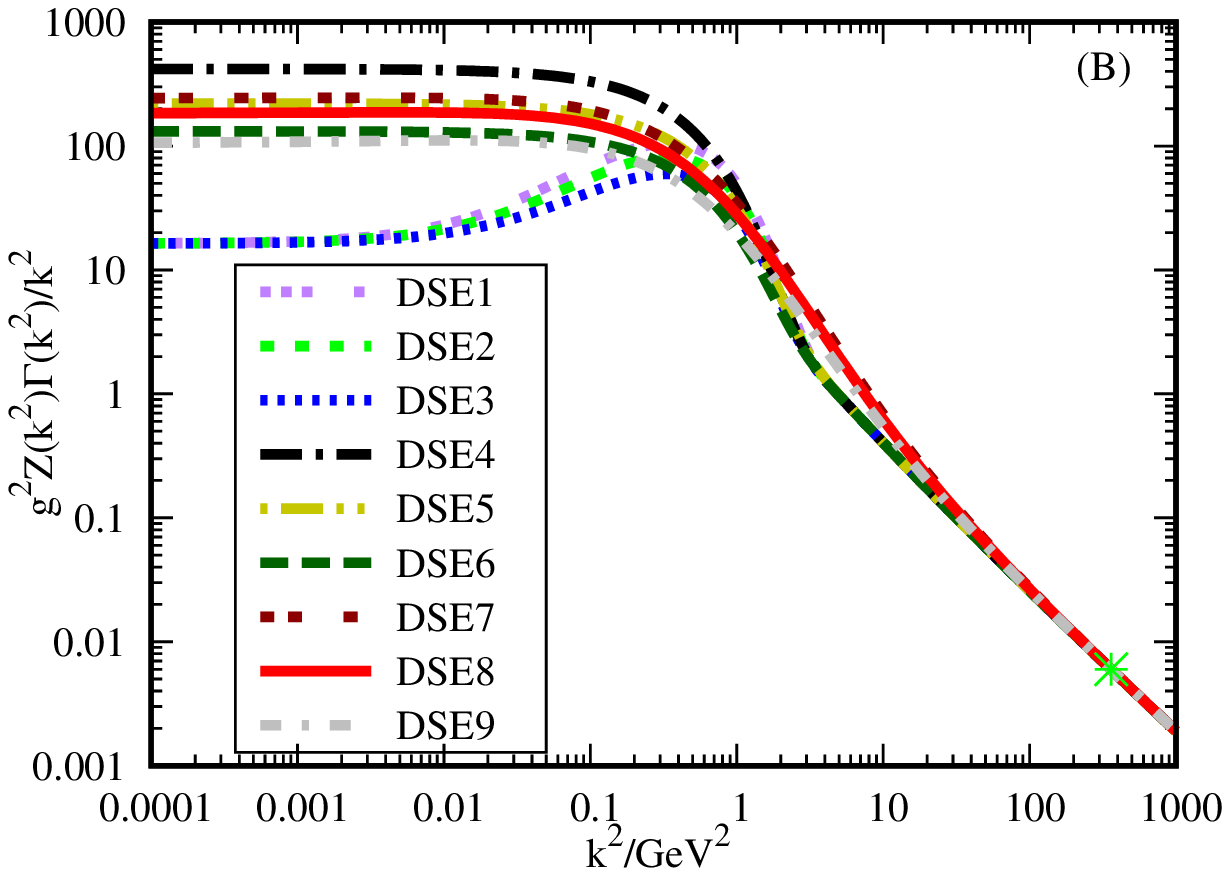}}
	\end{minipage}
}
\vspace*{0mm} \caption{\label{InputWinger-vertexdress}
\textcolor{black}{
The various models of the quark-gluon vertex dressing function $g^2\Gamma({k^2})$ (left panel) and the effective interaction ${\cal G}(k^2)/k^{2}=g^2Z^N_{}(k^{2}){\Gamma}(k^2)/k^{2}$ (right panel). See the text for details of the notation "DSE1",$...$, "DSE9" .}
}
\end{figure*}
\begin{figure*}[htp!]
 \centering
 \includegraphics[scale=0.95]{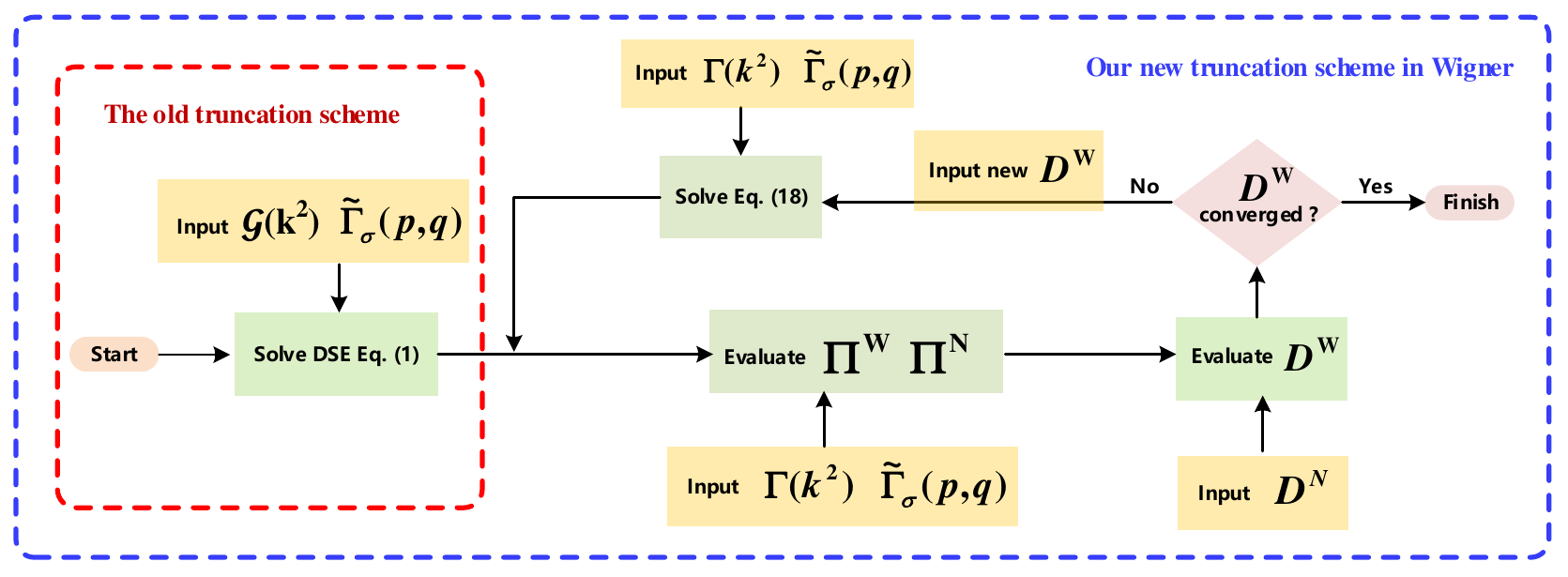}
 \caption{\label{Fig:workflow}%
\textcolor{black}{
Typical micro/macrocycle program flow for solving the coupled DSEs for quark propagator and gluon propagator. The old truncation scheme solving only the DSE for the quark propagator is shown in the small red flow box.
}
}
\end{figure*}

\section{Numerical Results}
\begin{figure*}[htp!]
	\centering
	
\subfigure
{
	\begin{minipage}{0.3\linewidth}
	\centering
	\centerline{\includegraphics[scale=0.652,angle=0]{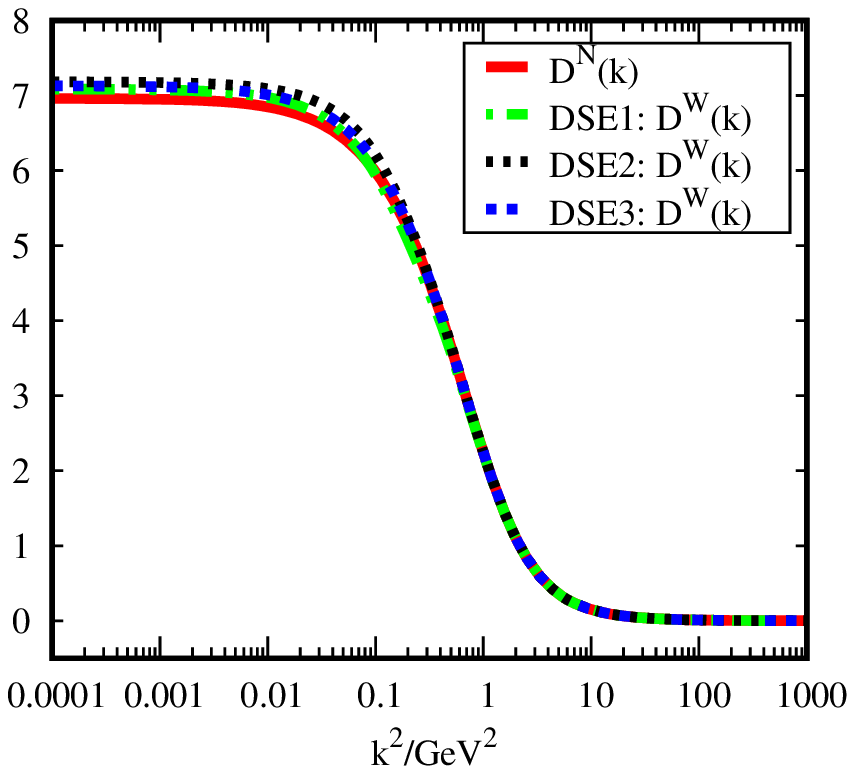}}
	\end{minipage}
}
\vspace{-0.2cm}
\subfigure
{
	\begin{minipage}{0.3\linewidth}
	\centering
	\centerline{\includegraphics[scale=0.652,angle=0]{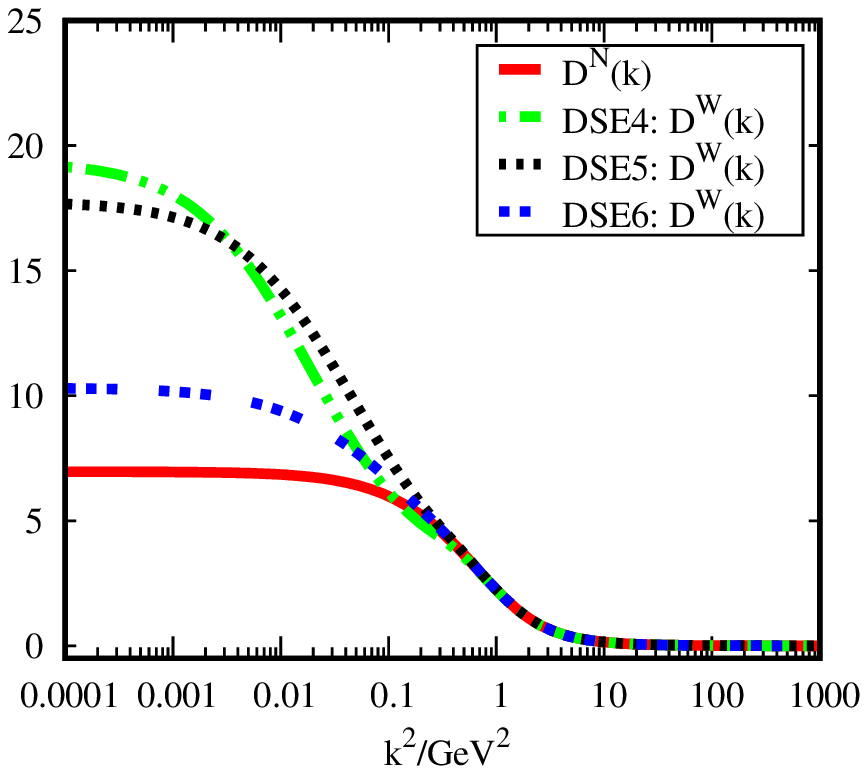}}
	\end{minipage}
}
\subfigure
{
	\begin{minipage}{0.3\linewidth}
	\centering
	\centerline{\includegraphics[scale=0.652,angle=0]{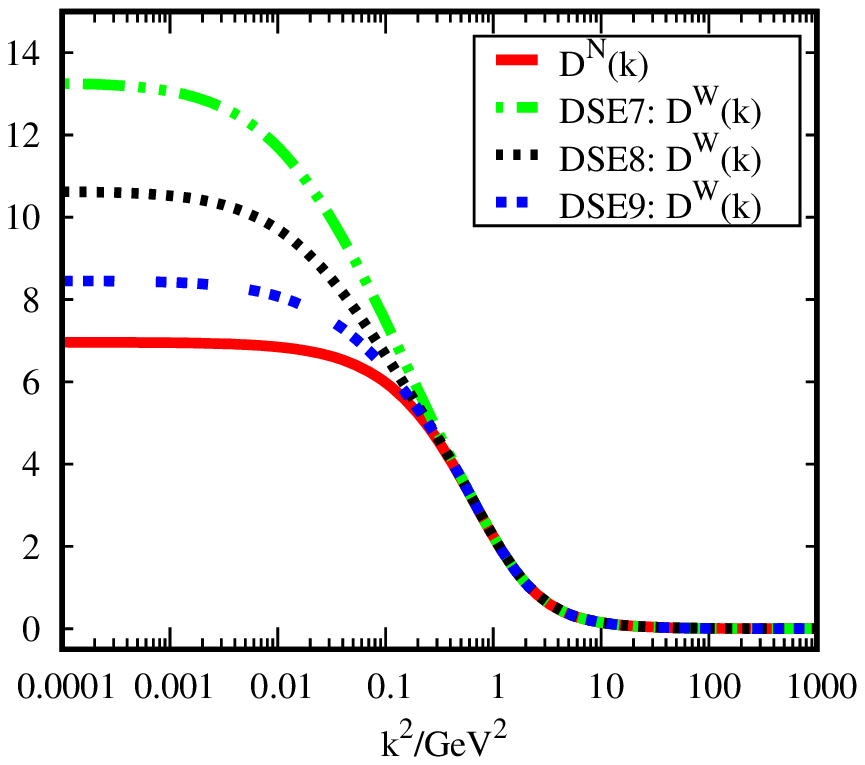}}
	\end{minipage}
}
\vspace{-0.25cm}

\subfigure
{
	\begin{minipage}{0.3\linewidth}
	\centering
	\centerline{\includegraphics[scale=0.652,angle=0]{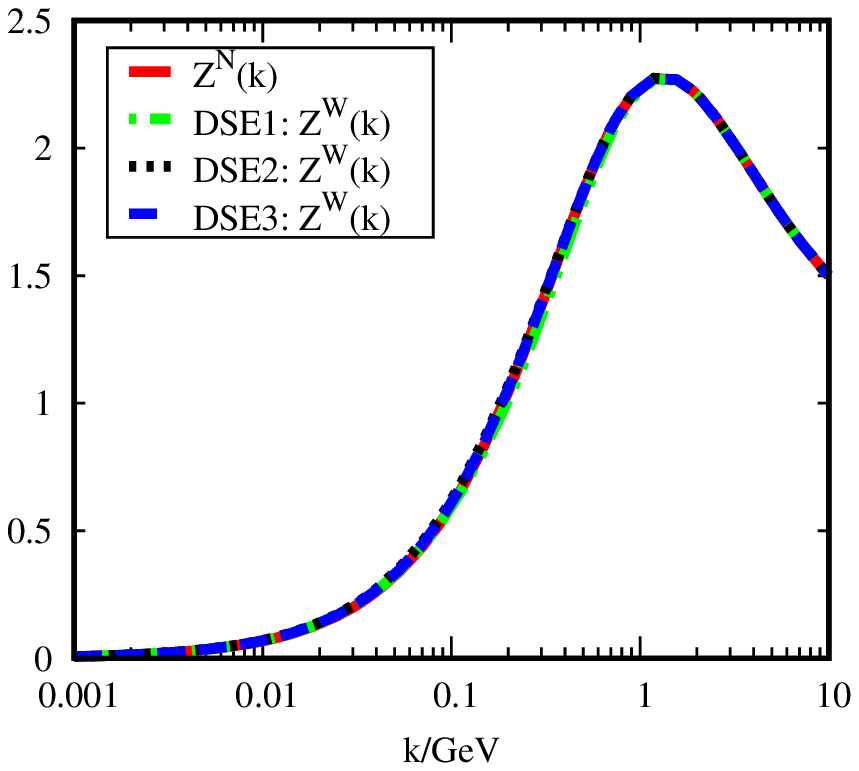}}
	\end{minipage}
}
\subfigure
{
	\begin{minipage}{0.3\linewidth}
	\centering
	\centerline{\includegraphics[scale=0.652,angle=0]{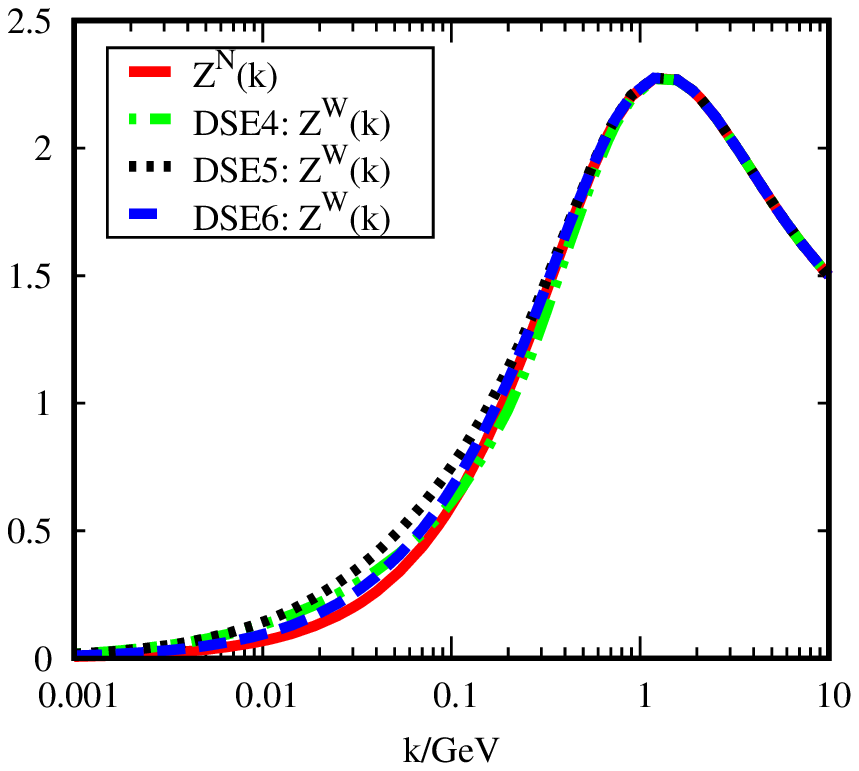}}
	\end{minipage}
}
\subfigure
{
	\begin{minipage}{0.3\linewidth}
	\centering
	\centerline{\includegraphics[scale=0.652,angle=0]{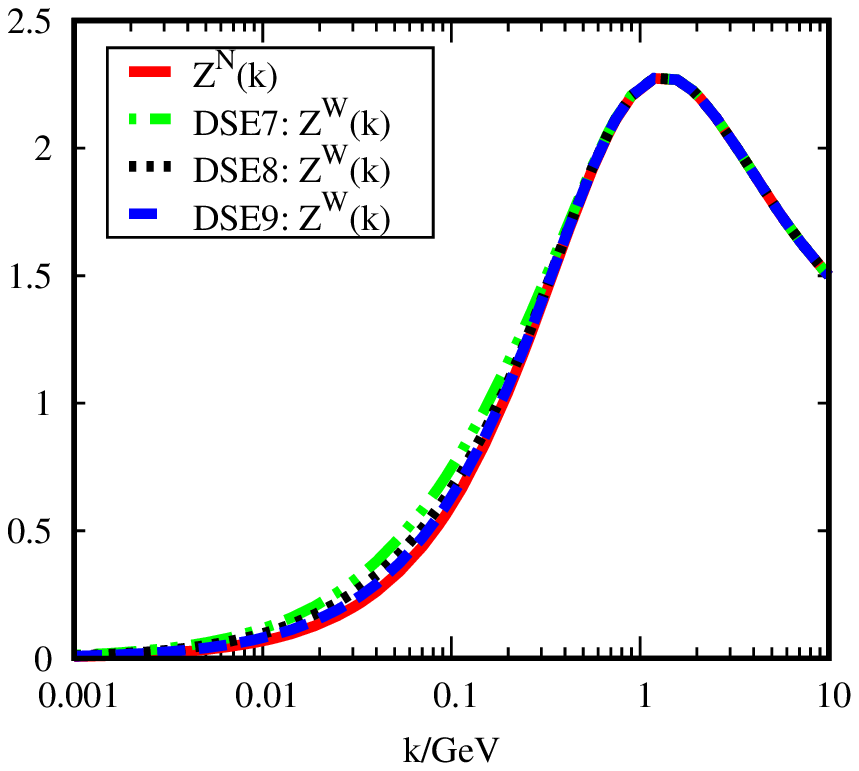}}
	\end{minipage}
}
\vspace{-0.25cm} 
\caption{\label{gluon-solution} The numerical results of the gluon propagator function (upper panels) and gluon dressing function (lower panels) with various models (see Table~\ref{tab:table4} and Table~\ref{tab:table5} for model details). The superscript 'N/W' represent the Nambu/Wigner phase respectively. }
\end{figure*}

\begin{figure*}[htp!]
	\centering
	\vspace{-0.2cm}
\subfigure
{
	\begin{minipage}{0.3\linewidth}
	\centering
	\centerline{\includegraphics[scale=0.652,angle=0]{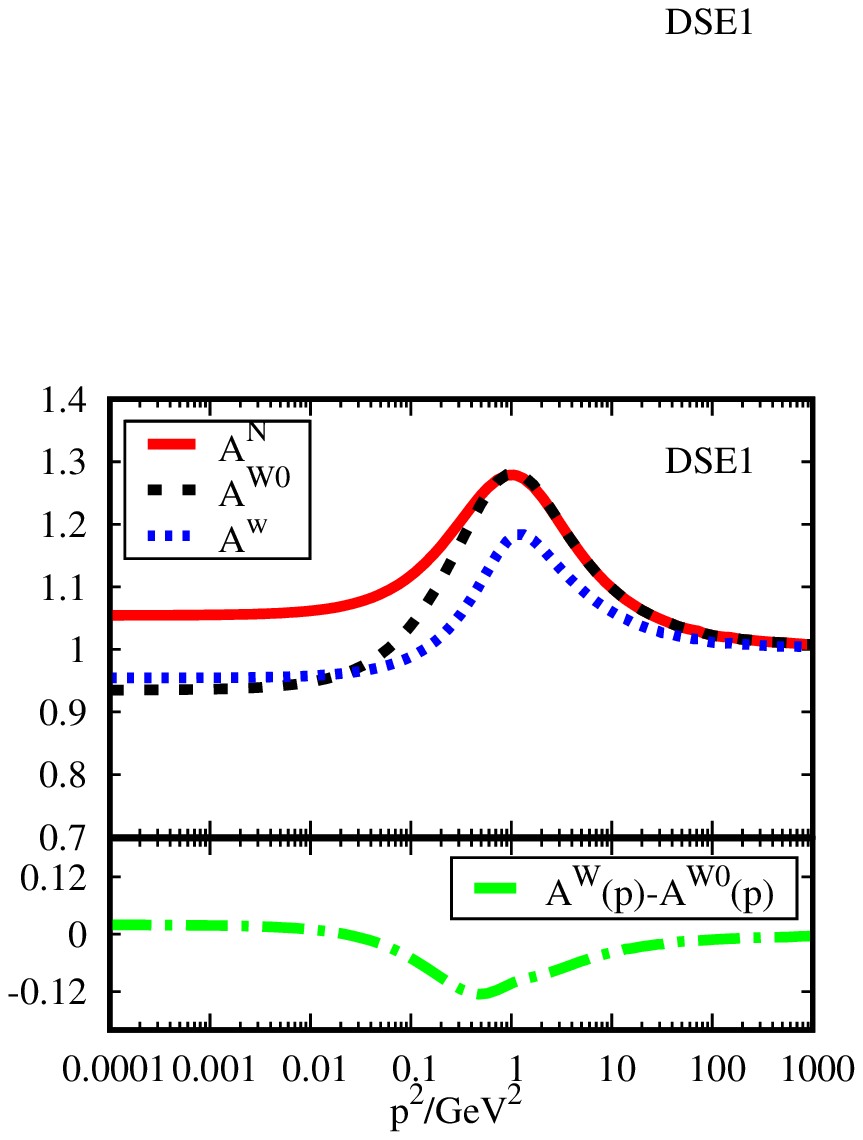}}
	\end{minipage}
}
\subfigure
{
	\begin{minipage}{0.3\linewidth}
	\centering
	\centerline{\includegraphics[scale=0.652,angle=0]{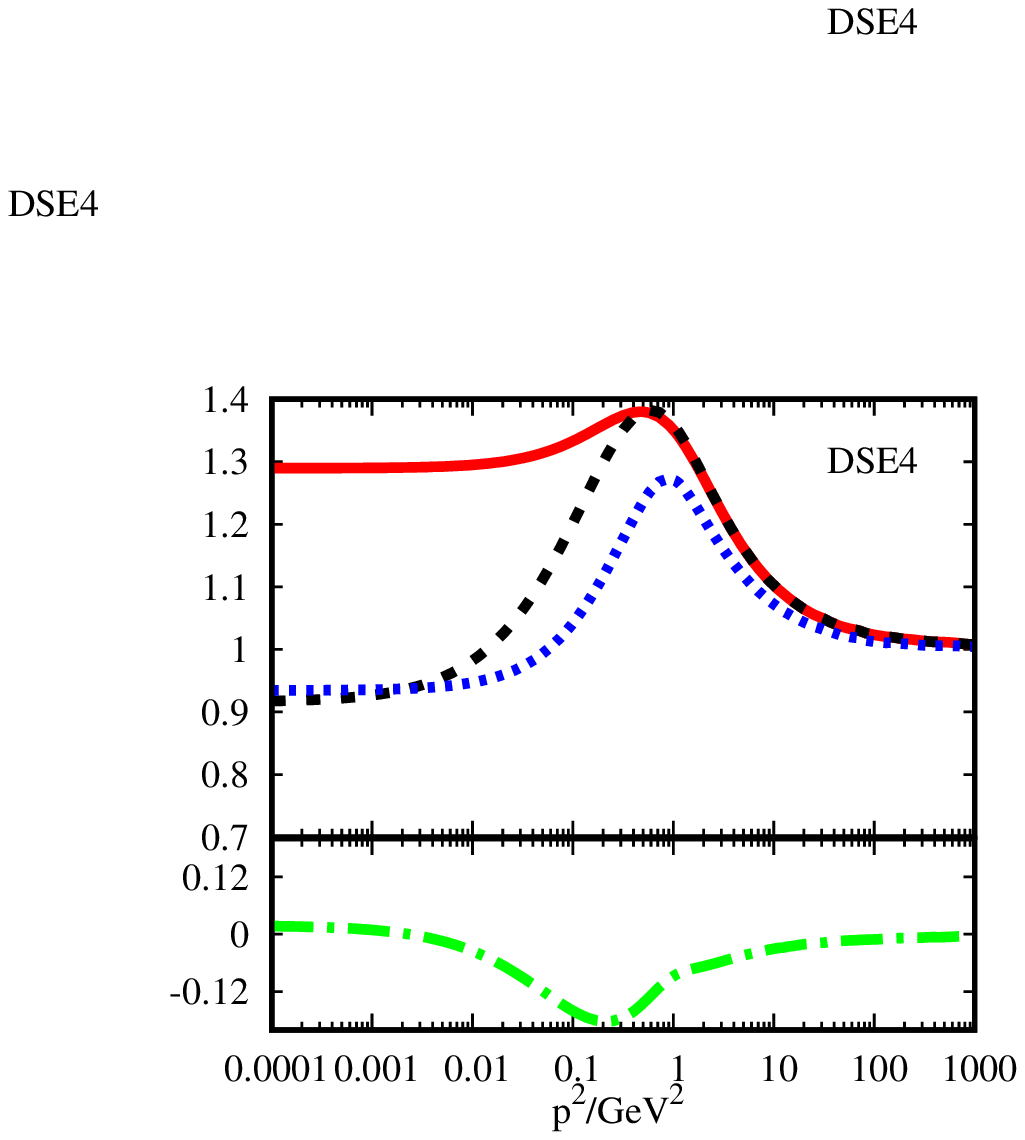}}
	\end{minipage}
}
\subfigure
{
	\begin{minipage}{0.3\linewidth}
	\centering
	\centerline{\includegraphics[scale=0.652,angle=0]{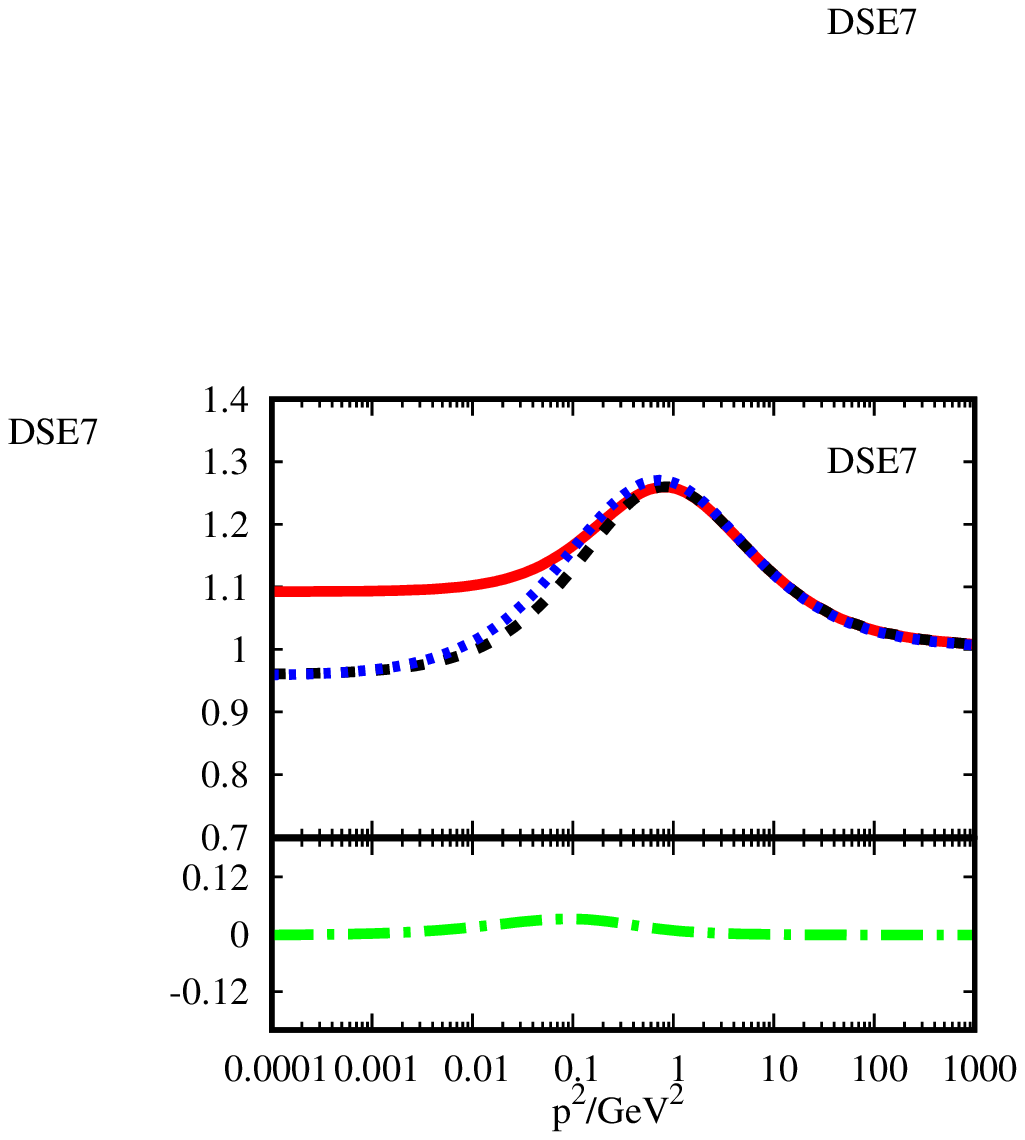}}
	\end{minipage}
}

\subfigure
{
	\begin{minipage}{0.3\linewidth}
	\centering
	\centerline{\includegraphics[scale=0.652,angle=0]{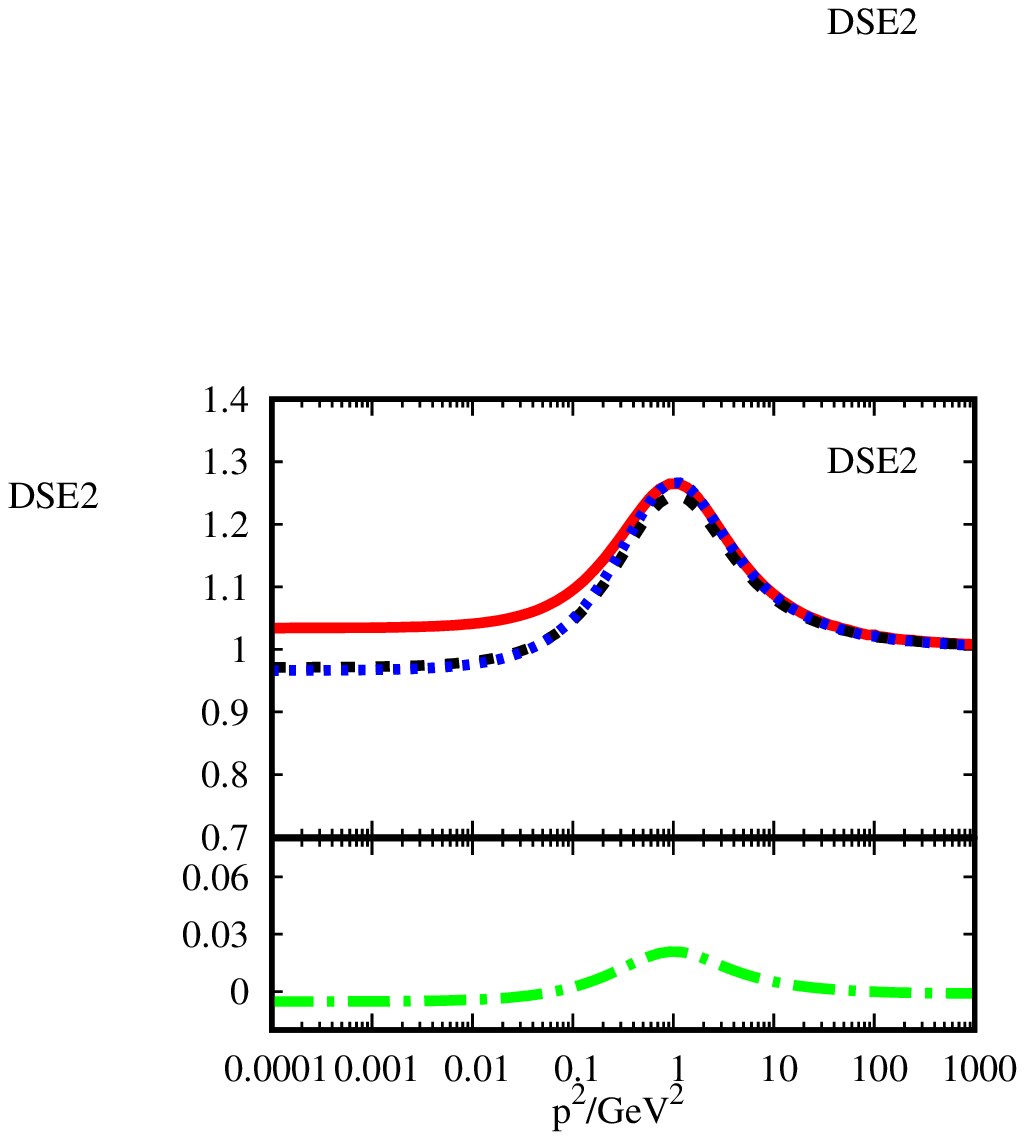}}
	\end{minipage}
}
\vspace{-0.2cm}
\subfigure
{
	\begin{minipage}{0.3\linewidth}
	\centering
	\centerline{\includegraphics[scale=0.652,angle=0]{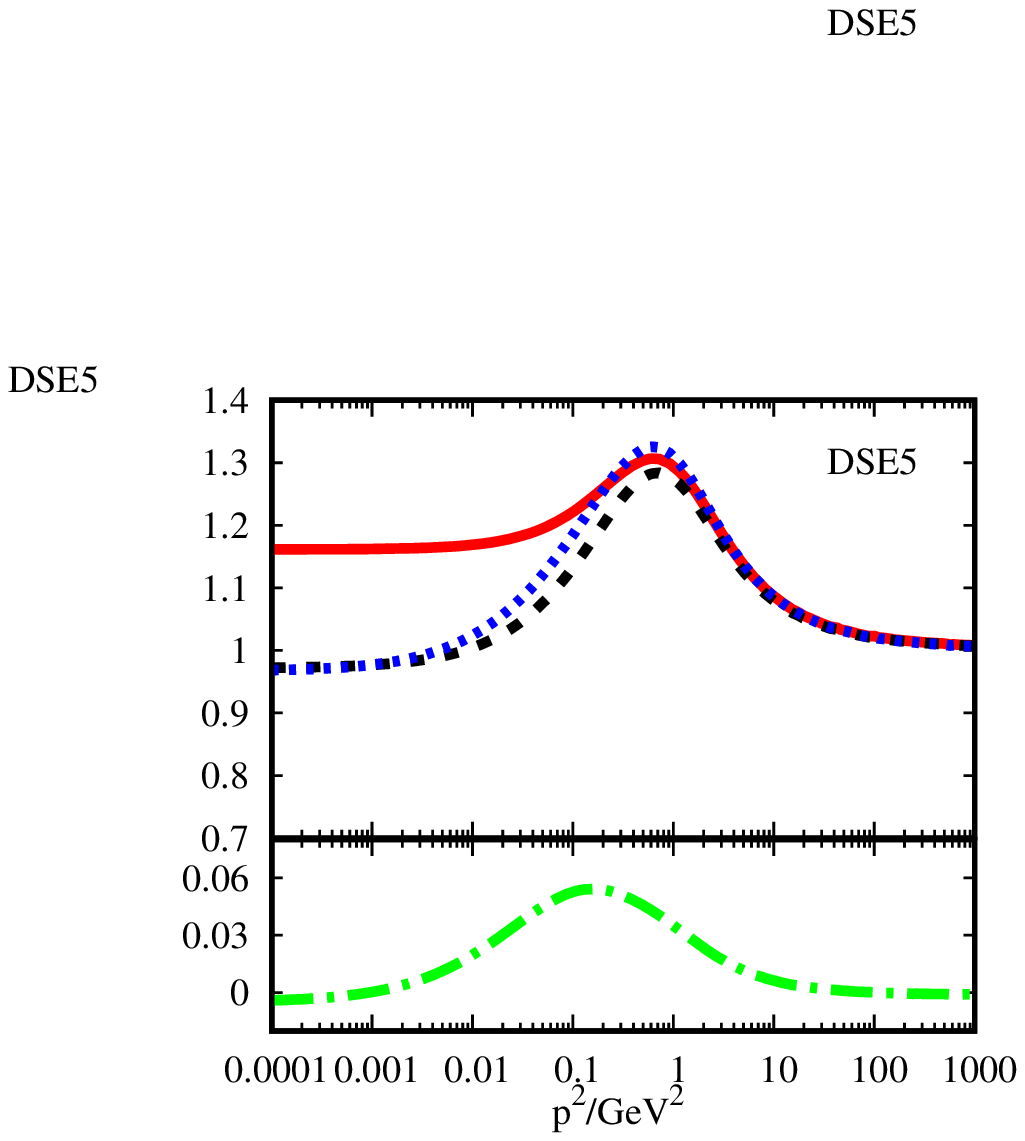}}
	\end{minipage}
}
\subfigure
{
	\begin{minipage}{0.3\linewidth}
	\centering
	\centerline{\includegraphics[scale=0.652,angle=0]{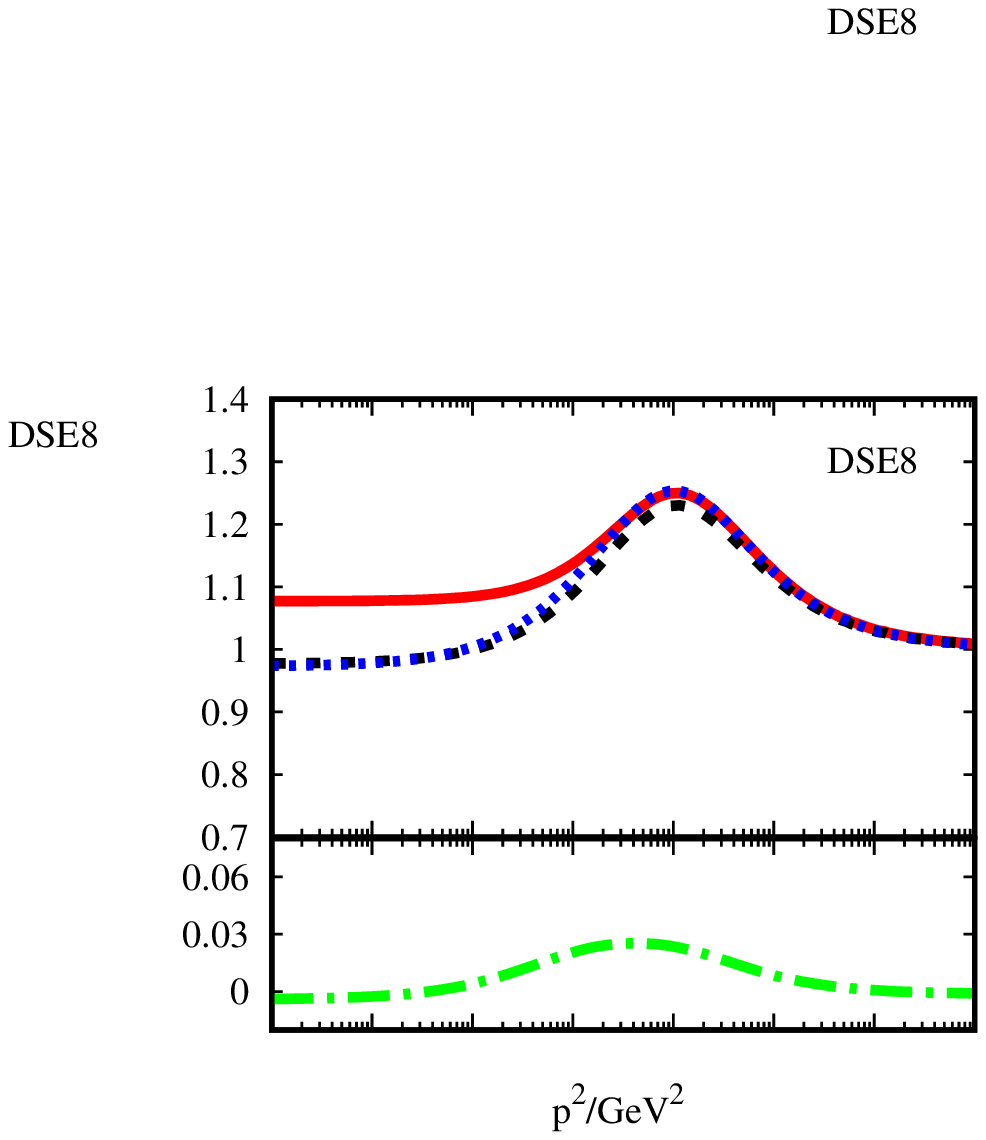}}
	\end{minipage}
}

\subfigure
{
	\begin{minipage}{0.3\linewidth}
	\centering
	\centerline{\includegraphics[scale=0.652,angle=0]{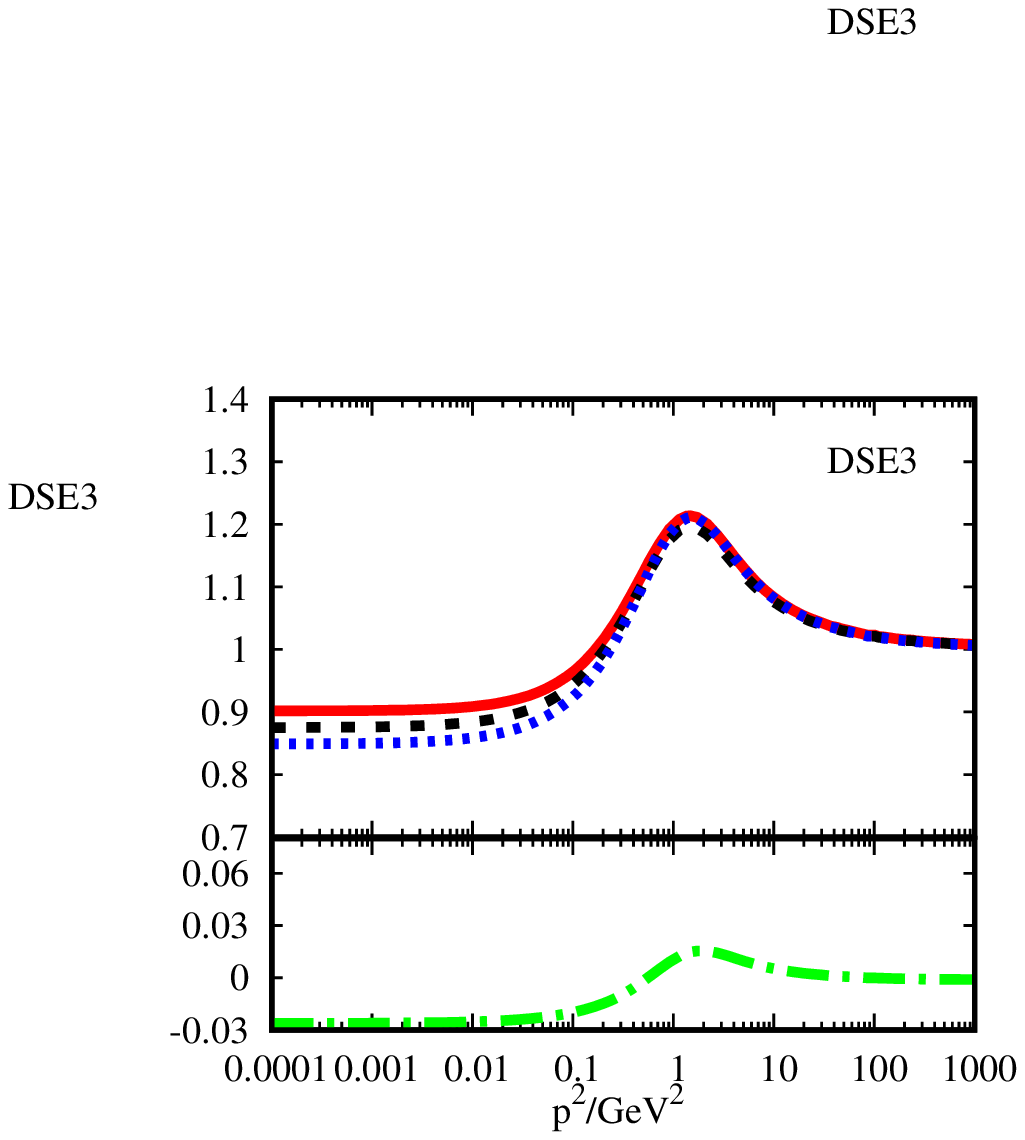}}
	\end{minipage}
}
\subfigure
{
	\begin{minipage}{0.3\linewidth}
	\centering
	\centerline{\includegraphics[scale=0.652,angle=0]{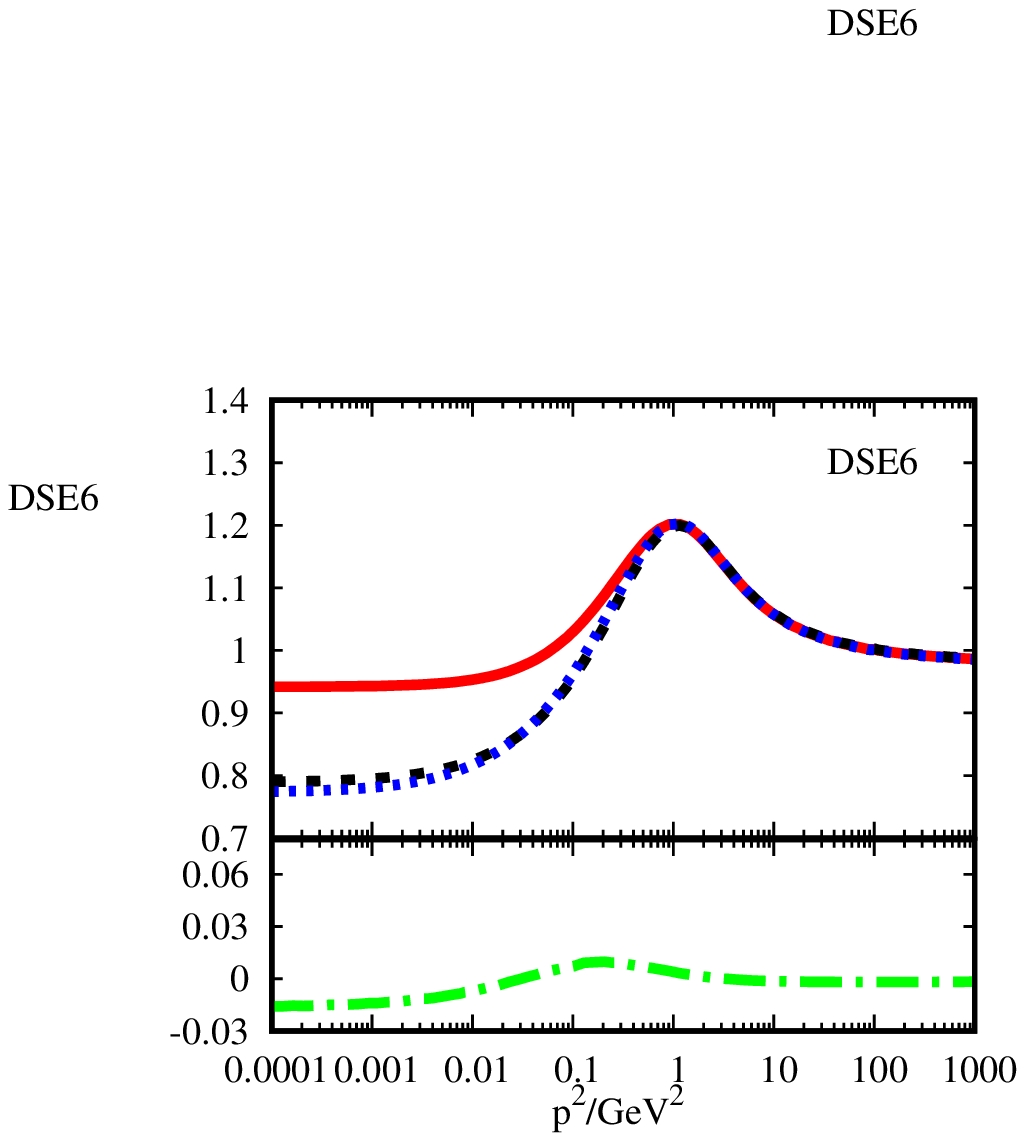}}
	\end{minipage}
}
\subfigure
{
	\begin{minipage}{0.3\linewidth}
	\centering
	\centerline{\includegraphics[scale=0.652,angle=0]{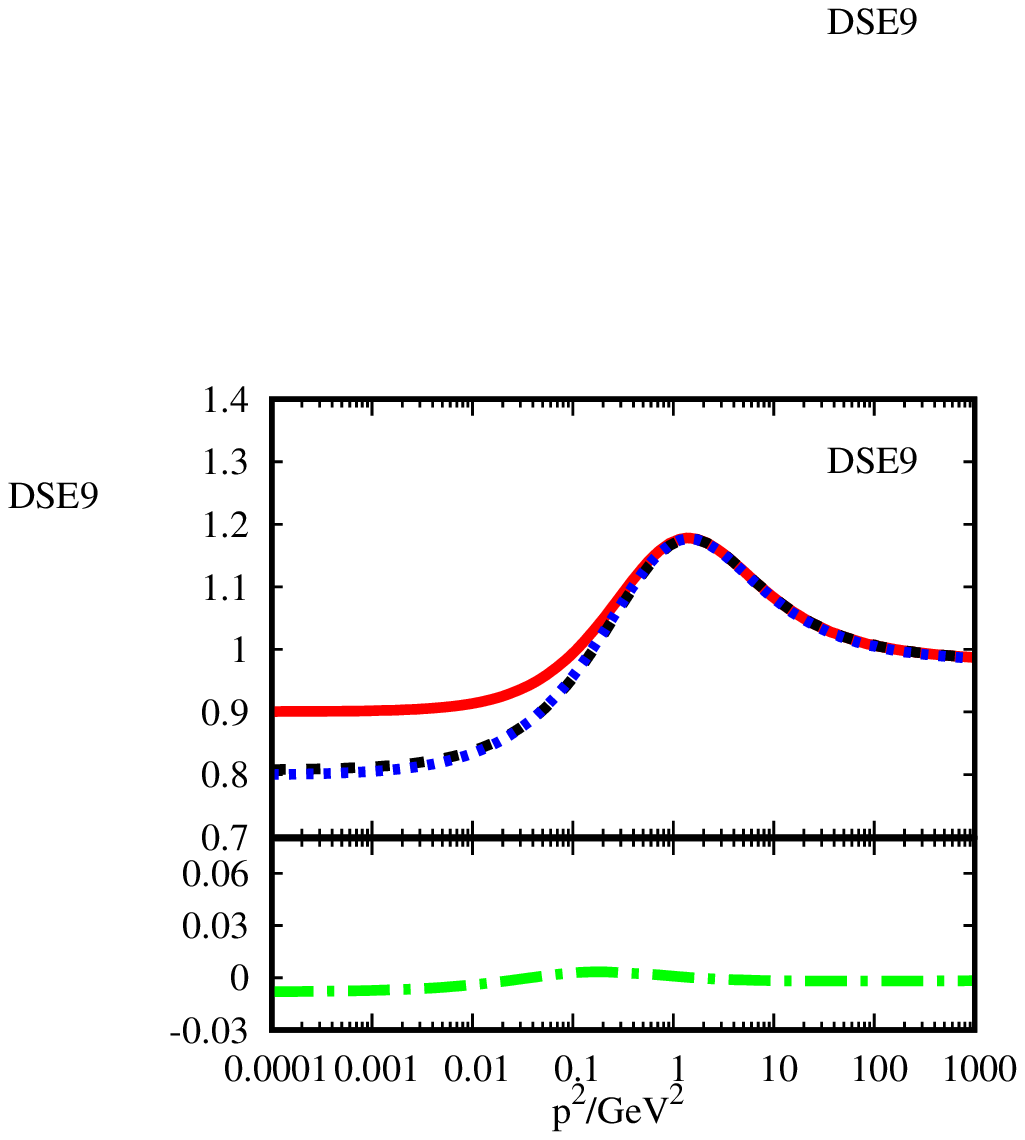}}
	\end{minipage}
}
\vspace{-0.25cm} \caption{\label{dse-solution}
The numerical results of the quark propagator function $A(p)$ with various models (see Table~\ref{tab:table4} and Table~\ref{tab:table5} for model details). The superscript ’N/W’ represents the Nambu/Wigner phase respectively, the superscript ’W0’ represents the Wigner solution obtained with the old truncation scheme. }
\end{figure*}

\begin{figure*}[htp!]
	\centering
	
\subfigure
{
	\vspace{-0.2cm}
	\begin{minipage}{8.2cm}
	\centering
	\centerline{\includegraphics[scale=0.652,angle=0]{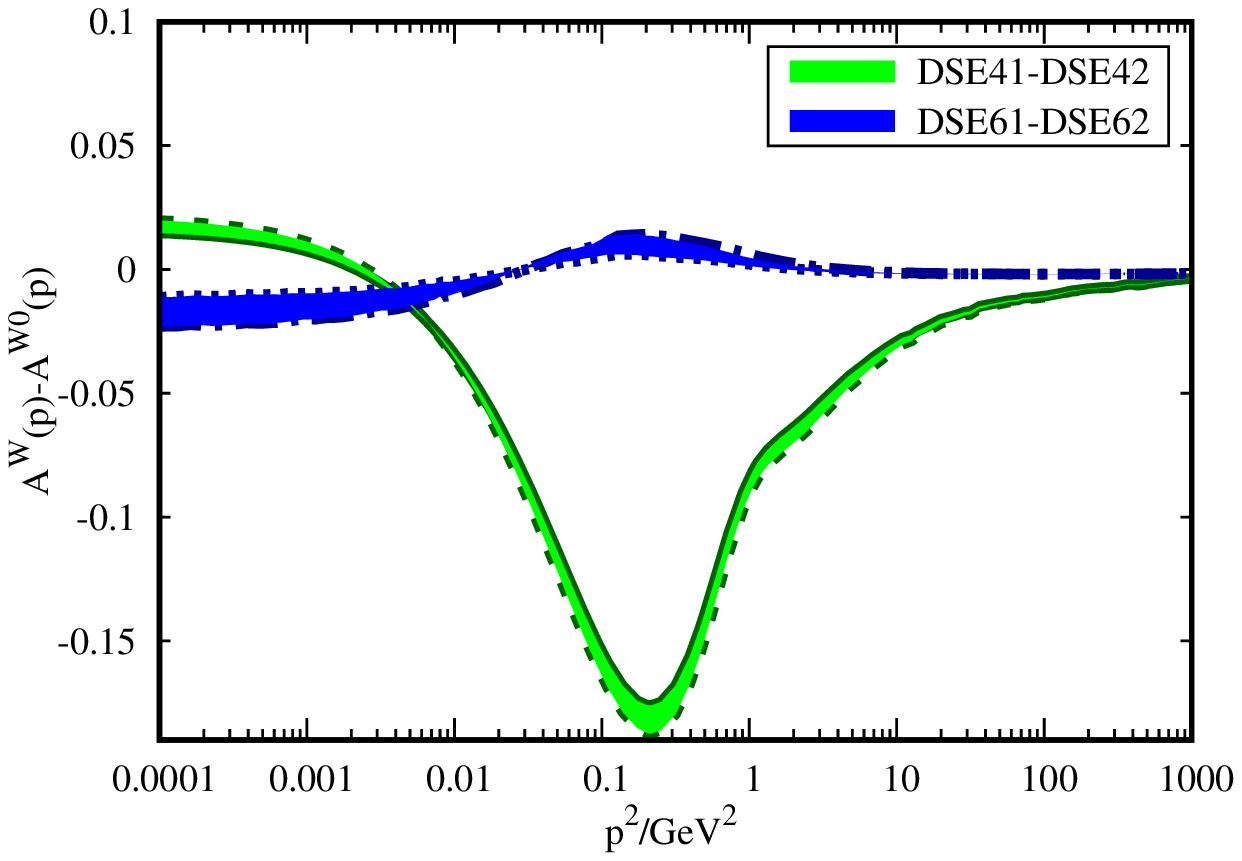}}
	\end{minipage}
}
\vspace{-0.2cm}
\subfigure
{
	 \vspace{-0.2cm}
	\begin{minipage}{8.2cm}
	\centering
	\centerline{\includegraphics[scale=0.652,angle=0]{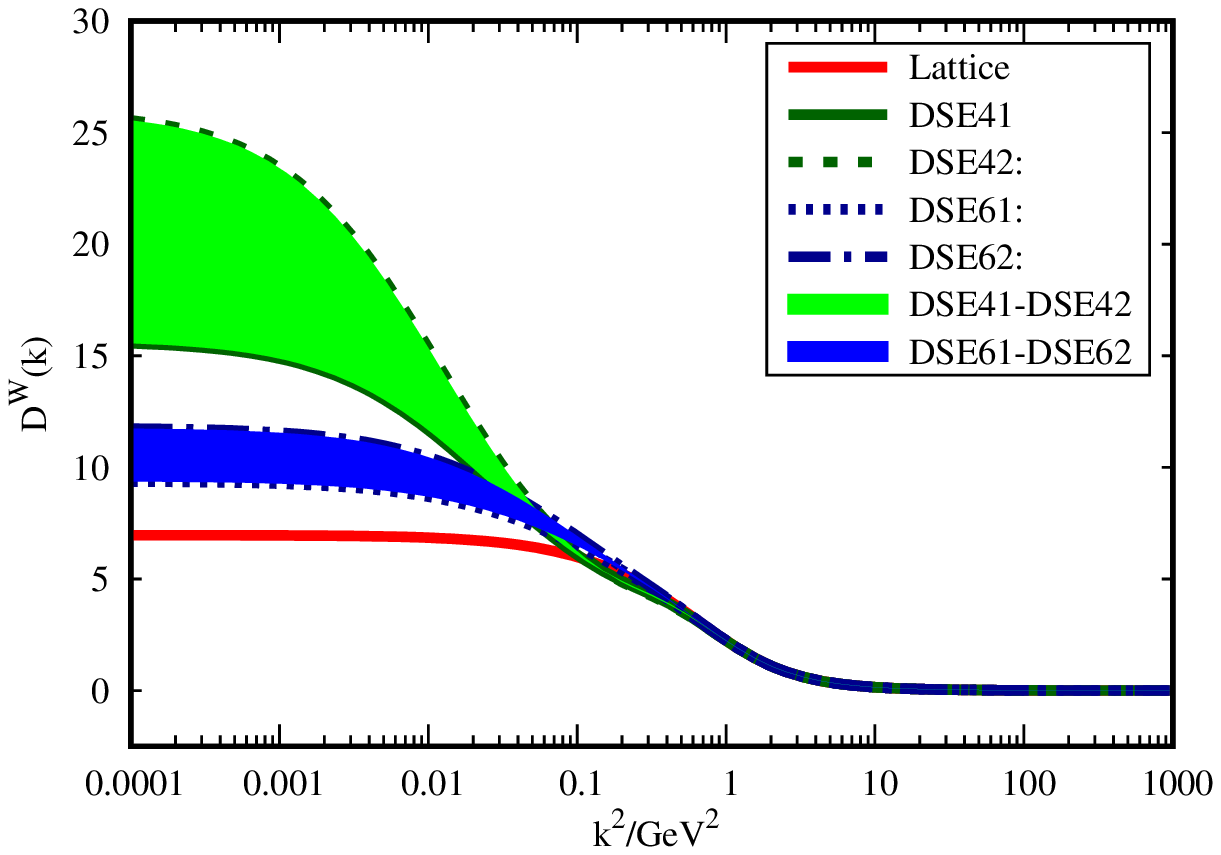}}
	\end{minipage}
}

\vspace*{0mm} \caption{\label{Fig:deltaWD} 
The dependence of the quark propagator function $A^W$ (left panel) and gluon propagator function $D^W$ (right panel) in the Wigner phase on parameters in models DSE4 and DSE6.
}
\end{figure*}

\quad 
To solve the coupled Eqs.~(\ref{Eq:qdseInWigner1}) and (\ref{Eq:qdseInWigner2}) numerically, we still need to fix the parameters. There are two parameters $\hat{D}$ and $\omega$ in the ansatz for the quark-gluon vertex dressing function ${{\Gamma}^{MT}(k^2)}$ and ${{\Gamma}^{QC}(k^2)}$(or the effective interaction ${\cal G}^{MT}(k^2)$, ${\cal G}^{QC}(k^2)$), which are usually fixed by fitting meson properties obtained by solving consistent \textcolor{black}{ Bethe-Salpeter equations~\cite{PhysRevC.84.042202,PhysRevD.103.074001}. } However, in the absence of such computation and for simplicity, we set $\hat{D}$ and $\omega$ by fitting the quark condensate in the chiral limit $-\langle \bar qq \rangle^{1/3}_{0}=N_{c}Z_{2} Z_{m} Tr[S(p)]$= $280\pm 10 {\rm MeV}$ and choosing $(\omega \hat{D})^{1/3}$ = const~\cite{Roberts-Hadron-1}. 
In the rainbow approximation, the typical value is chosen at $(\omega\hat{D})^{1/3}$ = 0.8 GeV~\cite{Qin2011}. With BC vertex, the typical value is chosen at $(\omega\hat{D})^{1/3}$ = 0.62 GeV~\cite{Chen2008}. Meanwhile, followed Ref.~\cite{Chen2015}, the typical value is chosen at $(\omega\hat{D})^{1/3}$ = 0.72 GeV with BC1 vertex.
The model and parameter setting are shown in Table.~\ref{tab:table4}. We marked the models as "DSE1",$...$, "DSE6", with the correspondence in the table.~\ref{tab:table4}. \textcolor{black}{
Similarly, the model and parameter setting with the vertex dressing function ${{\Gamma}^{CF}(k^2)}$ are shown in Table.~\ref{tab:table5}, marked correspondingly as "DSE7",$...$, "DSE9".
}

\textcolor{black}{The various models of the quark-gluon vertex dressing function $g^2\Gamma(k^2)$ and corresponding effective interaction ${\cal G}(k^2)=g^2 Z^{N}(k^{2}){\Gamma}(k^2)$ are shown in Fig.\ref{InputWinger-vertexdress}. Qualitatively, the vertex dressing function $g^2\Gamma^{MT}(k^2)$ are much suppressed in the deep infrared, while $g^2\Gamma^{MT}(k^2)$ and $g^2\Gamma^{CF}(k^2)$ go to finite constants in infrared. Quantitatively, with the same vertex structure, $g^2\Gamma^{CF}(k^2)$ in the intermidum momentum region, especially around a few $GeV$, are higher than $g^2\Gamma^{MT}(k^2)$ and $g^2\Gamma^{QC}(k^2)$, but lower than $g^2\Gamma^{QC}(k^2)$ in the deep infrared. For the same vertex dressing function model with various vertex structures, the strength of $g^2\Gamma(k^2)$ with the rainbow approximation is the highest, and it is lower with the BC1 vertex and lowest with BC vertex, for the later vertex structures providing more dressing effects. The effective interaction
${\cal G}(k^2)=g^2 Z^{N}(k^{2}){\Gamma}(k^2)$ in the right panel show similar properties. 
The vertex dressing functions and the effective interaction in all models coincide highly in ultraviolet,
with $g^2\Gamma(k^2)$=$g^2 Z^{N}(k^2)\Gamma(k^2)$=$4 \pi \alpha (\xi)$ at the renormalization
point $k^2={\xi}^2=(19GeV)^2$, which is marked as the star in the figure.}

\textcolor{black}{
The flow diagram for numerically solving the coupled Eqs.(\ref{Eq:qdseInWigner1}) and (\ref{Eq:qdseInWigner2}) is shown in Fig.~\ref{Fig:workflow}. We first solve the DSE Eq.~(\ref{Eq:quarkDSEdefine}) with input of Eq.~(\ref{KernelAnsatz}), obtaining the Nambu solution and the Wigner solution of the quark propagator.
Then we calculate the gluon propagator in the Wigner phase by adding the corrections of the quark loop to the gluon vacuum polarization. In the second line of Fig.~\ref{traunc_Quark_Gluon_DSE}, the first term of the right hand of the equation is the gluon propagator in Nambu phase, the second term is the vacuum polarisation calculated with the quark propagator in the Wigner phase and the third term is the vacuum polarisation with the quark propagator in the Nambu phase.
We then replace the gluon propagator in Eq.~(\ref{Eq:qdseInWigner1}) to re-calculate the quark propagator in the Wigner phase and iterate until obtaining the self-consistent solution of the coupled Eqs.(\ref{Eq:qdseInWigner1}) and (\ref{Eq:qdseInWigner2}).}
The coupled equations Eq.~(\ref{Eq:qdseInWigner1}) and Eq.~(\ref{Eq:qdseInWigner2}) are solved numerically using Gauss-Seidel relation, modified Levenberg-Marquardt minimization, and iteration. We find the iterations converged rapidly within 12 times.  

Fig.~\ref{gluon-solution} shows the gluon propagator function and gluon dressing function with various vertex structures and vertex dressing functions. \textcolor{black}{One can see some common features with all the models. The corrections in the ultraviolet are negligible and main corrections only appear in the infrared ($k^2 < 0.5GeV^2$).} In the left panels, we show the results with the MT model and various vertex structures. One can see only moderate corrections to the infrared gluon propagator in the winger phase, compared with that in the Nambu phase. With all three quark vertex models, there is only a little increase in the IR strength of the gluon propagator function. In the middle panels, we show the results with the QC model and various vertex structures. One can see the corrections to the infrared gluon propagator functions are much larger than those with the MT model and the results quantitatively depend strongly on the quark-gluon vertex structures. Especially in rainbow approximation and with the BC1 vertex, the infrared gluon propagator in the Wigner phase deviates largely from that in the Nambu phase. \textcolor{black}{In the right panels, we show the results with the CF model and various vertex structures. The corrections to the infrared gluon propagator function with the CF model are a little smaller than those with the QC model, but still much larger than those with the MT model. Comparing results with the same vertex dressing function model, the corrections with the BC vertex are smaller than those with the rainbow approximation and BC1 vertex.}
 
Fig.~\ref{dse-solution} shows the numerical results of the quark propagator functions with various vertex structures and vertex dressing functions. Due to the change of the gluon propagator compared with that in the old truncation scheme, the quark scalar function $A^{W}(p)$ in the Wigner phase is different from $A^{W0}(p)$, which is obtained with the old truncation scheme. The difference between $A^{W}(p)$ and $A^{W0}(p)$ are also shown in Fig.~\ref{dse-solution}. \textcolor{black}{Again, we find that all the corrections in the ultraviolet are negligible. However, it is different from the corrections on the gluon propagator, the corrections on the quark propagator appear both in the infrared and intermedium momentum region.} As to the model dependence, we can see the results quantitatively depend more on the quark-gluon vertex structure but less on the vertex dressing function. Remember that the deviation of $A(p)$ from the unit represents the interaction effects. \textcolor{black}{In the top panels, we can see that corrections $A^{W}(p)-A^{W0}(p)$ with rainbow approximation as well as the MT and QC models are large though the corrections on gluon propagator are small with the MT model and rainbow approximation. Furthermore, the corrections to the quark propagator function in all the other models are small, though the corrections to the gluon propagator function are large in models, e.g. DSE5 and DSE7.} 

\textcolor{black}{To investigate the parameter dependence of our results, we varied the fitted value of the chiral condensate. Take the example of the models DSE4 and DSE6, with the QC model for the vertex dressing function, as well as the rainbow approximation and BC vertex for the vertex structure, we show the corresponding parameters in Table.~\ref{tab:table4} and the results in Fig.~\ref{Fig:deltaWD}.} We can see that the parameter dependence of the corrections on the quark propagator in the Wigner phase is weak, while the parameter dependence of the gluon propagator corrections is strong, especially with the rainbow approximation. Quantitatively, the corrections to the gluon propagator with rainbow approximation are still much larger than that with the BC vertex.

\section{Summary}
\textcolor{black}{
The Wigner phase of QCD is important in studying the phase transition at high temperature/chemical potentials. In this work, we investigate the Wigner phase in the DSE framework. 
In the old truncation scheme, the Wigner phase is studied by solving only the DSE for the quark propagator. Models of gluon propagator and quark-gluon vertex are usually taken the same as in the Nambu phase. However, the feedback of quarks on the gluon propagator and quark-gluon vertex in the Wigner phase are usually neglected. 
Therefore we study the Wigner phase with an improved truncation scheme.  For the gluon propagator, we start from the $N_f$ = 2 unquenched results of lattice QCD in the Nambu phase and modify that in the Wigner phase by solving the coupled DSEs of the quark propagator and gluon propagator including the contribution from the quark loops. For the quark-gluon vertex, We employed three models for the vertex matrix structure: the rainbow approximation, the BC vertex and the BC1 vertex, and three models for the vertex dressing functions: the MT model, the QC model and the CF model.}

By solving the coupled equations, We obtain the gluon propagator and quark propagator in the Wigner phase and compare them with those obtained in the old truncation scheme. We find that the quark loop effects on the gluon propagator and quark propagator depend much on the quark-gluon vertex models. The corrections to the gluon propagator are small with the MT model and various vertex structures.
\textcolor{black}{
However, the corrections with the QC model and CF model are much larger, especially with the rainbow approximation and BC1 vertex. 
}
For the quark propagator, the corrections to the Wigner solution, i.e.  $A^{W}(p) - A^{W0}(p)$ is quite small with BC vertex, irrelevant to the dressing vertex function. However,  $A^{W}(p) - A^{W0}(p)$ is much larger in rainbow approximation with vertex dressing function $\Gamma^{MT}(k^2)$ and $\Gamma^{QC}(k^2)$. 

Herein, we should note, as discussed in previous chapters, that only moderate corrections in the Wigner phase are expected for self-consistency in our truncation scheme. If the corrections of the gluon propagator due to quark loops are large in the Wigner phase, then further corrections of the gluon propagator need to be included, e.g. the gluon loop corrections. Therefore, it looks much self-consistent only with the MT model for our present truncation scheme, but further investigation on the gluon loop corrections with the QC model or the CF model is needed. 

\textcolor{black}{
In this work, we employed some phenomenological models for the quark-gluon vertex. It is very important to go beyond this and employ more realistic results~\cite{Ayala:2012pb,PhysRevD.102.114518,PhysRevD.98.014002} on the quark-gluon vertex. We would improve our work with the more realistic quark-gluon vertex in future.}

To investigate phase transitions and quark matter in compact stars, it is needed to promote our present work to finite chemical potentials~\cite{li2010,Chen2012star,Chen2015,Chen2016,Wei2017,Luo2019}. The work is in progress.

\vspace{18pt}

\section*{Authors contributions}
Jing-Hui Huang and Huan Chen collected the literature and wrote the article. Guang-Jun Wang, Xiang-Yun Hu, and Qi Wang revised the article. Jing-Hui Huang designed the study. Xue-Ying Duan prepared figures and tables. All the authors were involved in the preparation of the manuscript. All the authors have read and approved the final manuscript.

\section*{Acknowledgements}
This study is financially supported by the National Science Foundation of China (No. 41630317), MOST Special Fund from the State Key Laboratory of Geological Processes and Mineral Resources, China University of Geosciences (No. MSFGPMR01-4), the National Key Research and Development Program of China (Grant No. 2018YFC1503705), and the Fundamental Research Funds for the Central Universities of Ministry of Education of China (Grant No. G1323519204).

\section*{Data Availability Statement}
This manuscript has no associated data or the data will not be deposited. Authors’comment: The processed data required to reproduce these findings cannot be shared at this time as the data also forms part of an ongoing study.
\bibliographystyle{ws-mpla}
\bibliography{dscs}

\begin{thebibliography}{10}

\bibitem{Klahn2016}
T.~Klahn, T.~Fischer and M.~Hempel, {\em Astrophys. J.} {\bf 836},  ~89
  (2017).

\bibitem{QCDPT-DSE12-4}
F.~Gao, J.~Chen, Y.-X. Liu, S.-X. Qin, C.~D. Roberts and S.~M. Schmidt, {\em
  Phys. Rev.} {\bf D93},   094019  (2016).

\bibitem{Li:2018tut}
C.-M. Li, P.-L. Yin and H.-S. Zong, {\em Phys. Rev.} {\bf D99},   076006
  (2019).

\bibitem{Barnafoldi2007}
G.~G. Barnafoldi and V.~Gogokhia, {\em J. Phys.} {\bf G37},   025003  (2010).

\bibitem{Cornwall1974}
J.~M. Cornwall, R.~Jackiw and E.~Tomboulis, {\em Phys. Rev.} {\bf D10}, 2428
  (1974).

\bibitem{bag2}
G.~G. Barnafoldi and V.~Gogokhia, {\em J. Phys.} {\bf G37},   025003  (2010).

\bibitem{bag4}
R.~Hofmann, T.~Gutsche, M.~Schumann and R.~D. Viollier, {\em Eur. Phys. J.}
  {\bf C16}, 677  (2000).

\bibitem{Masuda2014}
K.~Masuda, T.~Hatsuda and T.~Takatsuka, {\em PoS} {\bf Hadron2013},   169
  (2013).

\bibitem{Masuda2015}
K.~Masuda, T.~Hatsuda and T.~Takatsuka, {\em Eur. Phys. J.} {\bf A52},  ~65
  (2016).

\bibitem{Masuda2016}
K.~Masuda, T.~Hatsuda and T.~Takatsuka, {\em Nucl. Phys.} {\bf A956}, 817
  (2016).

\bibitem{DElia2019}
M.~D'Elia, F.~Negro, A.~Rucci and F.~Sanfilippo, {\em Phys. Rev.} {\bf D100},
  054504  (2019).

\bibitem{PhysRevD.99.014007}
D.~Horvati\ifmmode~\acute{c}\else \'{c}\fi{}, D.~Kekez and
  D.~Klabu\ifmmode~\check{c}\else \v{c}\fi{}ar, {\em Phys. Rev. D} {\bf 99},
  014007 (Jan 2019).

\bibitem{PhysRevLett.127.262301}
STAR Collaboration Collaboration, M.~S. e.~a. Abdallah, {\em Phys. Rev. Lett.}
  {\bf 127},   262301 (Dec 2021).

\bibitem{PhysRevD.94.054008}
R.~D. Pisarski and V.~V. Skokov, {\em Phys. Rev. D} {\bf 94},   054008 (Sep
  2016).

\bibitem{FISCHER2011438}
C.~S. Fischer, J.~Luecker and J.~A. Mueller, {\em Phys. Lett. B} {\bf 702}, 438
   (2011).

\bibitem{PhysRevD.101.054032}
W.-j. Fu, J.~M. Pawlowski and F.~Rennecke, {\em Phys. Rev. D} {\bf 101},
  054032 (Mar 2020).

\bibitem{Golanbari_2020}
T.~Golanbari, T.~Haddad, A.~Mohammadi, M.~A. Rasheed and K.~Saaidi, {\em Chin.
  Phys. C} {\bf 44},   083109 (aug 2020).

\bibitem{Bazavov2019}
A.~Bazavov, F.~Karsch, S.~Mukherjee, P.~Petreczky and U.~Collaboration, {\em
  Eur. Phys. J. A} {\bf 55},   194  (2019).

\bibitem{Guenther2021}
J.~N. Guenther, {\em Eur. Phys. J. A} {\bf 57},   136  (2021).

\bibitem{PhysRevD.100.024061}
A.~Tsokaros, M.~Ruiz, V.~Paschalidis, S.~L. Shapiro and K.~b.~o.
  Ury\ifmmode~\bar{u}\else \={u}\fi{}, {\em Phys. Rev. D} {\bf 100},   024061
  (Jul 2019).

\bibitem{ROBERTS2021103883}
C.~D. Roberts, D.~G. Richards, T.~Horn and L.~Chang, {\em Prog. Part. Nucl.
  Phys.} {\bf 120},   103883  (2021).

\bibitem{Roberts-Hadron-7}
S.-x. Qin, C.~D. Roberts and S.~M. Schmidt, {\em Few Body Syst.} {\bf 60},  ~26
   (2019).

\bibitem{EICHMANN2012234}
G.~Eichmann, {\em Prog. Part. Nucl. Phys.} {\bf 67}, 234  (2012).

\bibitem{CLOET20141}
I.~C. Cloët and C.~D. Roberts, {\em Prog. Part. Nucl. Phys.} {\bf 77}, 1
  (2014).

\bibitem{HUBER20201}
M.~Q. Huber, {\em Physics Reports} {\bf 879}, 1  (2020), Nonperturbative
  properties of Yang-Mills theories.

\bibitem{QCDPT-DSE11}
S.-x. Qin, L.~Chang, H.~Chen, Y.-x. Liu and C.~D. Roberts, {\em Phys. Rev.
  Lett.} {\bf 106},   172301  (2011).

\bibitem{QCDPT-DSE12-5}
F.~Gao and Y.-x. Liu, {\em Phys. Rev.} {\bf D94},   094030  (2016).

\bibitem{QCDPT-DSE23}
C.~S. Fischer, {\em Prog. Part. Nucl. Phys.} {\bf 105}, 1  (2019).

\bibitem{Chen2015}
H.~Chen, J.~B. Wei, M.~Baldo, G.~F. Burgio and H.~J. Schulze, {\em Phys. Rev.}
  {\bf D91},   105002  (2015).

\bibitem{PhysRevD.103.103003}
T.-T. Sun, Z.-Y. Zheng, H.~Chen, G.~F. Burgio and H.-J. Schulze, {\em Phys.
  Rev. D} {\bf 103},   103003 (May 2021).

\bibitem{Bai2021}
Z.~Bai and Y.-x. Liu, {\em Eur. Phys. J. C} {\bf 81},   612  (2021).

\bibitem{Sternbeck2010}
A.~Sternbeck and L.~von Smekal, {\em Eur. Phys. J. C} {\bf 68}, 487  (2010).

\bibitem{PhysRevD.86.114513}
O.~Oliveira and P.~J. Silva, {\em Phys. Rev. D} {\bf 86},   114513 (Dec 2012).

\bibitem{PhysRevD.102.114518}
A.~F. Falc\~ao, O.~Oliveira and P.~J. Silva, {\em Phys. Rev. D} {\bf 102},
  114518 (Dec 2020).

\bibitem{DSE-1-1}
C.~D. Roberts and A.~G. Williams, {\em Prog. Part. Nucl. Phys.} {\bf 33}, 477
  (1994).

\bibitem{DSE-1-2}
C.~D. Roberts and S.~M. Schmidt, {\em Prog. Part. Nucl. Phys.} {\bf 45}, S1
  (2000).

\bibitem{DSE-1-3}
A.~Bashir, L.~Chang, I.~C. Cloet, B.~El-Bennich, Y.-X. Liu, C.~D. Roberts and
  P.~C. Tandy, {\em Commun. Theor. Phys.} {\bf 58}, 79  (2012).

\bibitem{HE200932}
M.~He, F.~Hu, W.-M. Sun and H.-S. Zong, {\em Physics Letters B} {\bf 675}, 32
  (2009).

\bibitem{YUAN200669}
W.~Yuan, H.~Chen and Y.~xin Liu, {\em Physics Letters B} {\bf 637}, 69  (2006).

\bibitem{Papavassiliou:2014qva}
J.~Papavassiliou, {\em PoS} {\bf QCD-TNT-III},   029  (2013).

\bibitem{Meyers:2014iwa}
J.~Meyers and E.~S. Swanson, {\em Phys. Rev.} {\bf D90},   045037  (2014).

\bibitem{Lowdon:2018mbn}
P.~Lowdon, {\em Phys. Lett.} {\bf B786}, 399  (2018).

\bibitem{Ayala:2012pb}
A.~Ayala, A.~Bashir, D.~Binosi, M.~Cristoforetti and J.~Rodriguez-Quintero,
  {\em Phys. Rev.} {\bf D86},   074512  (2012).

\bibitem{PhysRevD.98.014002}
A.~C. Aguilar, J.~C. Cardona, M.~N. Ferreira and J.~Papavassiliou, {\em Phys.
  Rev. D} {\bf 98},   014002 (Jul 2018).

\bibitem{PhysRevD.81.065003}
M.~Q. Huber, R.~Alkofer and S.~P. Sorella, {\em Phys. Rev. D} {\bf 81},
  065003 (Mar 2010).

\bibitem{PhysRevD.100.056001}
C.~Tang, F.~Gao and Y.-x. Liu, {\em Phys. Rev. D} {\bf 100},   056001 (Sep
  2019).

\bibitem{DUDAL2018351}
D.~Dudal, O.~Oliveira and P.~J. Silva, {\em Ann. Phys.} {\bf 397}, 351  (2018).

\bibitem{Oliveira2019}
O.~Oliveira, W.~de~Paula, T.~Frederico and J.~P. B.~C. de~Melo, {\em Eur. Phys.
  J. C} {\bf 79},   116  (2019).

\bibitem{PhysRevD.103.114515}
A.~K\ifmmode \imath \else \i \fi{}z\ifmmode \imath \else~\i \fi{}lers\"u,
  O.~Oliveira, P.~J. Silva, J.-I. Skullerud and A.~Sternbeck, {\em Phys. Rev.
  D} {\bf 103},   114515 (Jun 2021).

\bibitem{Williams2015}
R.~Williams, {\em Eur. Phys. J.} {\bf A51},  ~57  (2015).

\bibitem{PhysRevD.97.054006}
A.~K. Cyrol, M.~Mitter, J.~M. Pawlowski and N.~Strodthoff, {\em Phys. Rev. D}
  {\bf 97},   054006 (Mar 2018).

\bibitem{Maris1999}
P.~Maris and P.~C. Tandy, {\em Phys. Rev.} {\bf C60},   055214  (1999).

\bibitem{Qin2011}
S.-x. Qin, L.~Chang, Y.-x. Liu, C.~D. Roberts and D.~J. Wilson, {\em Phys.
  Rev.} {\bf C84},   042202  (2011).

\bibitem{PhysRevC.84.042202}
S.-x. Qin, L.~Chang, Y.-x. Liu, C.~D. Roberts and D.~J. Wilson, {\em Phys. Rev.
  C} {\bf 84},   042202 (Oct 2011).

\bibitem{PhysRevD.103.074001}
L.~Chang and M.~Ding, {\em Phys. Rev. D} {\bf 103},   074001 (Apr 2021).

\bibitem{PhysRevD.80.074029}
C.~S. Fischer and J.~A. Mueller, {\em Phys. Rev. D} {\bf 80},   074029 (Oct
  2009).

\bibitem{Fischer2010fx}
C.~S. Fischer, A.~Maas and J.~A. Muller, {\em Eur. Phys. J.} {\bf C68}, 165
  (2010).

\bibitem{Fischer2012vc}
C.~S. Fischer and J.~Luecker, {\em Phys. Lett.} {\bf B718}, 1036  (2013).

\bibitem{PhysRevD.101.014016}
R.~Contant and M.~Q. Huber, {\em Phys. Rev. D} {\bf 101},   014016 (Jan 2020).

\bibitem{Muller2013}
D.~M{\"{u}}ller, M.~Buballa and J.~Wambach, {\em Eur. Phys. J. A} {\bf 49},
  ~96  (2013).

\bibitem{BC19811}
J.~S. Ball and T.-W. Chiu, {\em Phys. Rev.} {\bf D22},   2542  (1980).

\bibitem{PhysRevD.99.074013}
L.~Liu, L.~Chang and Y.-x. Liu, {\em Phys. Rev. D} {\bf 99},   074013 (Apr
  2019).

\bibitem{BC1981}
J.~S. Ball and T.-W. Chiu, {\em Phys. Rev.} {\bf D22},   2550  (1980).

\bibitem{Zong2002}
H.-s. Zong, J.-l. Ping, W.-m. Sun, F.~Wang and C.-h. Chang, {\em Int. J. Mod.
  Phys.} {\bf A21}, 3387  (2006).

\bibitem{Chen2008}
H.~Chen, W.~Yuan, L.~Chang, Y.-X. Liu, T.~Klahn and C.~D. Roberts, {\em Phys.
  Rev.} {\bf D78},   116015  (2008).

\bibitem{Roberts-Hadron-1}
C.~D. Roberts, {\em Prog. Part. Nucl. Phys.} {\bf 61}, 50  (2008).

\bibitem{li2010}
H.~Li, X.-L. Luo and H.-S. Zong, {\em Phys. Rev.} {\bf D82},   065017  (2010).

\bibitem{Chen2012star}
H.~Chen, M.~Baldo, G.~F. Burgio and H.~J. Schulze, {\em Phys. Rev.} {\bf D86},
   045006  (2012).

\bibitem{Chen2016}
H.~Chen, J.~B. Wei and H.~J. Schulze, {\em Eur. Phys. J.} {\bf A52},   291
  (2016).

\bibitem{Wei2017}
J.~B. Wei, H.~Chen, G.~F. Burgio and H.~J. Schulze, {\em Phys. Rev.} {\bf D96},
    043008  (2017).

\bibitem{Luo2019}
Z.~H. Luo, J.~B. Wei, G.~Chen, H.~Chen and H.~J. Schulze, {\em Mod. Phys.
  Lett.} {\bf A34},   1950202  (2019).

\end{thebibliography}

\end{sloppypar}

\end{document}